\RequirePackage[l2tabu,orthodox]{nag}
\RequirePackage{pdftexcmds}
\RequirePackage{pdf14}

\documentclass[acmsmall]{acmart}
\usepackage{rotating}
\usepackage{todonotes} 
\usepackage{svg}
\usepackage[utf8]{inputenc}
\usepackage{graphicx}
\usepackage{subfig}
\usepackage{multirow}
\usepackage{tablefootnote}
\usepackage{lscape}
\usepackage{array}
\usepackage{tikz}

\begin{document}
	\acmJournal{CSUR} 
	\title{A Comprehensive Survey on Parallelization and Elasticity in Stream Processing}
	
	\author{Henriette Röger}
	\orcid{0000-0001-7139-6135}
	\affiliation{University of Stuttgart - Institute of Parallel and Distributed Systems}
	\email{henriette.roeger<at>ipvs.uni-stuttgart.de}

	\author{Ruben Mayer}
	\orcid{0000-0001-9870-7466}
	\affiliation{Technical University of Munich}
	\email{mayerr<at>in.tum.de}

 \copyrightyear{2019}
\acmYear{2019} 
\acmJournal{CSUR}

	
	\begin{abstract} 

Stream Processing (SP) has evolved as the leading paradigm to process and gain value from the high volume of streaming data produced e.g. in the domain of the Internet of Things. An SP system is a middleware that deploys a network of operators between data sources, such as sensors, and the consuming applications. SP systems typically face intense and highly dynamic data streams. Parallelization and elasticity enables SP systems to process these streams with continuously high quality of service. The current research landscape provides a broad spectrum of methods for parallelization and elasticity in SP. Each method makes specific assumptions and focuses on particular aspects of the problem. However, the literature lacks a comprehensive overview and categorization of the state of the art in SP parallelization and elasticity, which is necessary to consolidate the state of the research and to plan future research directions on this basis. Therefore, in this survey, we study the literature and develop a classification of current methods for both parallelization and elasticity in SP systems. 
  
	\end{abstract}
	
	\maketitle
	
	\begin{tikzpicture}
	
	\tikzset{curveinscope/.style={every path/.style={draw=white, double distance=1pt, line width=2pt, double=red, text=red}}}
	
	\begin{scope}[overlay]
	\node[text width=18cm] at ([yshift=-9cm]current page.south) {\color{red}(c) Owner 2019. This is the authors' version of the work. It is posted here for your personal use only. \newline Not for redistribution. \newline The definitive version will be published in ACM Computing Surveys};
	\end{scope}
	\end{tikzpicture}

\section{Introduction}

With the surge of the Internet of Things and digitalization in all areas of life, the volume of digital data available raises tremendously. A large share of this data is produced as continuous data streams. Stream Processing (SP) systems have been established as a middleware to process these streams to gain valuable insights from the data. 
For stepwise processing of the data streams, SP systems span a network of \emph{operators}---the operator graph. 
To ensure high throughput and low latency with the massive amount of data, SP systems need to parallelize processing. 
This parallelism comes with two major challenges: 
First, \textit{how} to parallelize the processing in SP operators. SP systems require mechanism to increase the level of parallelization, which is especially hard for stateful operators that require the state to be partitioned onto different CPU cores of a multi-core server or even different processing nodes in a shared-nothing cloud-based infrastructure. Frequent state synchronization must not hamper parallel processing, while the processing results have to remain consistent. Research proposes different approaches for parallel, stateless and stateful SP. They differ in assumptions about the operator functions and state externalization mechanisms an SP system supports. This led to the development of a broad range of parallelization approaches tackling different problem cases.

Second, how to continuously \textit{adapt} the level of parallelization when the conditions of the SP operators, e.g. the workload or resources available, change at runtime. 
On the one hand, an SP system always needs enough resources to process the input data streams with a satisfying quality of service (QoS), e.g. latency or throughput. On the other hand, continuous provisioning of computing resources for peak workloads wastes resources at off-peak hours. Thus, an \emph{elastic} SP system scales its resources according to the current need. Cloud computing provides on-demand resources to realize such elasticity \cite{armbrust2010view}. The pay-as-you-go business model of cloud computing allows to cut costs by dynamically adapting the resource reservations to the needs of the SP system. It is challenging to strive the right balance between resource over-provisioning---which is costly, but is robust to workload fluctuations---and on-demand scaling---which is cheap, but is vulnerable to sudden workload peaks. To this end, academia and industry developed elasticity methods. Again, they differ in their optimization objectives and assumptions about the operator parallelization model employed, the target system architecture, state management as well as timing and methodology.

While there are many works that propose methods and solutions for specific parallelization and elasticity problems in SP systems, there is a severe lack of overview, comparison, and classification of these methods. When we investigated these topics in more depth, we found more than 40 papers that propose methods for SP parallelization, and more than 25 papers that propose methods for SP elasticity. An even higher number of papers addresses related problems such as placement, scheduling and migration of SP operators. Every year, dozens of new publications appear in the major conferences and journals of the field.
 Furthermore, the authors of those papers are located in different sub-communities of the stream processing domain.
While they work on the same topics, they have a different view on the problems in SP parallelization and elasticity. Hence, there is an urgent need for a broad investigation, classification and comparison of the state of the art in methods for SP parallelization and elasticity.

\subsection{Complementary Surveys}

Cugola and Margara \cite{cugola_processing_2012} presented a general overview of SP systems, languages and concepts, but did not take into account parallelization and elasticity methods. In their survey, Hirzel et al. \cite{hirzel_catalog_2014} compared different general concepts to optimize SP operators; however, the authors did not investigate elasticity. 
Heinze et al. \cite{Heinze:2014:CDS:2611286.2611309} in their tutorial provided a broad overview of SP systems for cloud environments, but did not particularly focus on parallelization and elasticity. Mencagli and De Matteis  \cite{matteis_parallel_2017} investigated parallelization patterns for window-based SP operators. They did neither take into account other parallelization strategies nor discussed elasticity methods. Flouris et al. \cite{flouris_issues_nodate} discussed issues in SP systems that are executed in cloud environments. The authors discussed query representations, event selection strategies, probabilistic event streams, eager and lazy detection approaches, optimization with query rewriting, and memory management. However, they only briefly mentioned parallelization and elasticity.  
Basanta-Val et al. \cite{7953583} presented patterns to optimize real-time SP systems. These patterns cover operator decomposition and fusion, data parallelization, operator placement, and load shedding. The authors included a theoretical analysis of capabilities and overhead of those patterns. Their focus is on mathematical theory rather than a comprehensive study of specific parallelization and elasticity methods. Many elastic SP systems apply methods from control theory to adapt the parallelization degree of the operators. 
Shevtsov et al.~\cite{7929422} provide a systematic literature review on control-theoretical software adaptation that goes beyond SP systems. It can be read as a complement of this survey to get a larger view on adaptive software beyond SP systems. 
In a recent article, Assuncao et al. \cite{de2017resource} investigated parallel SP systems with a strong focus on infrastructure and architecture, details on how to implement SP and descriptions of open-source SP frameworks as well as SP in cloud environments. Their discussion of elasticity approaches however lacks a fine grained classification e.g. on parallelization strategy, timing, provided QoS and the methodology focus that is needed to understand the broad range of elasticity concepts. Again, it can be read as a complement of this survey with details how to realize parallel, elastic SP and a stronger focus on frameworks and SP in cloud computing.
We conclude that even though surveys for many aspects of SP systems are available, there is a need for a comprehensive study that navigates through available methods for parallelization and elasticity in SP systems to continously  deliver high QoS.

\subsection{Our Contributions}
In this article, we provide a broad analysis of properties relevant for parallel and elastic SP and introduce the methods for parallelization and elasticity. We further summarize and classify the parallelization and elasticity approaches in SP systems. This includes a discussion of research gaps and trends in the field. 
In addition, we discuss issues that are related to parallelization and elasticity, such as placement, scheduling and migration. 
These contributions should be helpful both to researchers as well as practitioners to assess the applicability of the state of the art methods to concrete scenarios. Further, they serve as a basis for future research.  
\subsection{Structure of the Survey}

We structured this survey as follows: In Section \ref{sec:fundamentals}, we discuss fundamentals of parallel SP systems and introduce the main methods for parallelization and elasticity. In Section \ref{sec:categorizationChapter}, we classify published work for parallel and elastic SP. Finally, we discuss related topics in Section~\ref{sec:related} and conclude the survey in Section~\ref{sec:conclusion}.



\section{General System Model and Classification}
\label{sec:fundamentals}

In this section, we briefly introduce a generic SP system model, properties of SP systems related to parallelization and parallelization and elasticity techniques.

\label{sec:categorizationSP}

\subsection{General System Model}
\label{sec:generalModel}
\begin{figure}[]
	\includegraphics[width=0.6\textwidth]{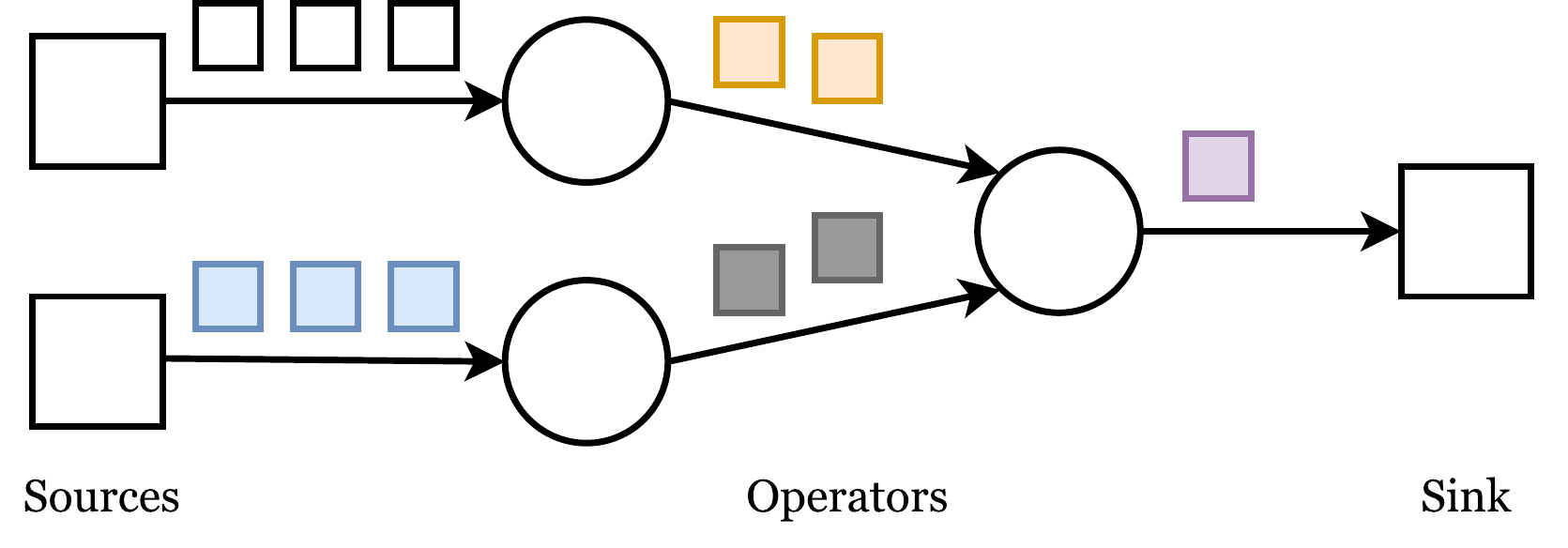}
	\caption{Schematic of an SP System. Sources emit streams of data items to be processed by a set of operators. A sink consumes the output data stream.}
	\label{fig:CEPOverview}
\end{figure}

An SP system processes data streams as the data arrives. It aggregates, filters and analyzes the data items and thus gains fast insights, reactions to observed situations and higher level information. Examples are continuous trend analysis of Twitter feeds, automatic stock trading, fraud detection and traffic monitoring. 
An SP systems core is the directed, acyclic operator graph that processes and forwards the input data in streams, also called its topology. Also to the topology belong the data sources that emit data items in streams into the graph and the sinks that consume the output. In Fig. \ref{fig:CEPOverview}, we provide a schematic of an exemplary SP system. Data sources are the rectangles on the left, operators the circles, connected by edges that represent the flow of data streams. The rectangle on the right depicts the sink. 
In a \textit{distributed} SP system, the operators run on multiple processing nodes that are connected via a communication network. 

As SP systems are designed to analyze data streams online, \textit{low latency} and \textit{high throughput} are the primary quality of service (\textit{QoS}) goals for SP systems and the major focus of parallelization and elasticity strategies. If a system does not meet its QoS goal, penalty fees may arise or the system, e.g. low latency alarm systems, becomes less useful. Secondary QoS goals, for example balanced load, node utilization, and fault tolerance, are often set to support low latency and high throughput.

\subsection{Parallel Stream Processing}
\label{sec:parallelizationClasses}

If the arrival rate of data items exceeds an operator's processing rate, the operator's input queue grows and induces queuing latency for the data items. Additionally, back-pressure might throttle operators that need to wait until the bottleneck downstream operator processes its input queue. The SP system's QoS goals might be jeopardized. \textit{Parallel SP} reduces queuing latencies and increases throughput as it processes multiple data items simultaneously instead of sequentially. SP systems can parallelize their processing in the multiple ways that we describe in the following. We start with a discussion of those SP system properties that influence the systems parallelization potential and continue with the parallelization methods. 

\subsection{Properties of Parallel SP Systems}
In this section, we introduce those properties of SP systems that influence the SP systems parallelization potential and are necessary for the discussion of approaches in Section \ref{sec:parallelization}.

\subsubsection{Type of SP System}
The type of the SP system determines what operations the system supports.
We distinguish General SP systems (GP) and the more specialized CEP systems (CEP): 

\emph{General SP systems.}
General SP systems apply \textit{continuous operations} on streams of data items. Each operator either produces an output stream of result data items (that itself can be the input stream of another operator) or make the result available to other applications, e.g., by writing them into a data storage or forwarding them to consuming sinks.
In general SP systems, we include Data Stream Management Systems (DSMS) that apply \textit{continuous queries} on data streams. A continuous query is a --- usually relational--- query continuously applied on a changing set of data like a data stream as opposed to traditional database applications that apply a changing set of queries on a fixed data set.
Classical examples of non-parallel DSMS systems are Aurora~\cite{Abadi:2003:ANM:950481.950485} and TelegraphCQ~\cite{DBLP:conf/cidr/ChandrasekaranDFHHKMRRS03}. Modern GP systems are for example Apache Storm~\cite{ApacheStorm} and Apache Flink~\cite{Carbone2015ApacheFS}. 

\emph{CEP.} 
CEP systems are SP systems dedicated to detect patterns of events and thereof derive higher level information. The patterns represent a complex \emph{situations of interest}, e.g. the pattern  ``\textit{Smoke} and \textit{high temperature''} represents \textit{``Fire''}. The input streams of CEP systems consist of \textit{events} triggered by observations of the surrounding world. The operators of CEP systems search for those \textit{sequences of events} in the input streams that fulfill the patterns. If an operator detects a pattern, it emits an output event (sometimes referred to as a \emph{complex event}, e.g. ``\textit{Fire}'').  When parallelizing CEP systems, it is necessary to divide the input stream onto parallel instances so that patterns will still be detected.
Common applications of CEP pattern detection are automatic stock trading \cite{balkesen_rip:_2013, mayer_spectre}, financial fraud detection~\cite{Alevizos:2017:EFP:3093742.3093920, Poppe:2017:CET:3035918.3035947} and traffic monitoring~\cite{mayer_predictable_2015}. 
A well-established open source CEP system is, for instance, Esper~\cite{Esper}. Besides that, there are general purpose SP systems that provide CEP functionality as a library, e.g., Apache Flink~\cite{Carbone2015ApacheFS}.

\subsubsection{Programming Model}
\label{sec:programmingmodel}
The \textit{programming model} defines if programmers use declarative \emph{queries} or an explicit \emph{imperative} implementation to specify the operations or patterns of an SP system.
Parallelization solutions for declarative systems might not work for imperative systems and vice versa. In addition, the programming model impacts the data model of the SP system. 

\emph{Declarative.} A declarative query follows syntax and semantics of a specific \emph{query language}. There are different query languages for CEP and for GP systems that differ in their expressiveness, i.e., which kind of queries can be specified in the language. For CEP systems, query languages such as Snoop \cite{chakravarthy_snoop:_1994}, SASE \cite{wu_high-performance_2006}, EPL \cite{Esper}, or TESLA \cite{cugola_tesla:_2010}, have been proposed.
Prominent DSMS-languages are Continuous Query Language (CQL)~\cite{Arasu:2006:CCQ:1146461.1146463} and (SPL), a query language for IBM's System S \cite{SystemS}. 
Declarative SP systems automatically deploy an operator graph that implements the query. This automatic deployment step can include optimizations, e.g., fusion or splitting of operators or enabling multiple queries to share operators, c.f. Section \ref{sec:efficientExecution}. Declarative SP systems usually require a structured data model for input data items. This data model is defined within the query.  
The query can then refer to this structure. A common meta model for the structured data model is the relational model.

\emph{Imperative.} A large group of SP systems follows the imperative programming model. It requires a programmatic specification of the operator graph. This includes both the specification of the operators themselves, e.g., by implementing an API of the SP system, and the specification of the topology of the operator graph, i.e., how the operators are connected. Imperative programming increases expressiveness as the definition of operations is not limited by a declarative language. The widely used open source frameworks Storm, Heron, and Flink follow an imperative programming model \cite{ApacheStorm}, \cite{ApacheFlink}, \cite{kulkarni2015twitter}.  
A shortcoming of the imperative model is that due to its less structured nature, automatic optimizations of the operator graph are harder to achieve as in the declarative model. 
Further, programmatic operators are black boxes to the SP system. The SP system lacks information about the operators internals and cannot exploit them for parallelization.
Imperative SP systems give the programmers more freedom for the structure of data items. While these items might still have defined header fields, the payload can be arbitrary.

It is possible to combine imperative and declarative programming. For example, declaratively defined operators are assembled imperatively into an operator graph \cite{tang2013autopipelining}.
The operator graph as the set of operators and their connections is termed the ``logical plan" of an SP system.
The assignment of this logical plan to a target infrastructure is referred to as the ``physical query plan''

\subsubsection{Sub-Stream Processing}
Operators usually process the input stream sub-stream wise. They extract sub streams based on \textit{keys} or \textit{windows}. 
\textit{Key-based} extraction groups data items by a the value of a key each data item needs to provide, leading to a sub stream per key value. E.g. in automatic stock trading, a possible key is the stocks class of business, where the trading SP system individually analyzes technology stocks and finance stocks \cite{hirzel_partition_2012}.

\textit{Window-based} extraction builds sub streams  according to a \textit{window policy}. A window-policy defines a scope and a slide. The scope describes the window size which, among others, can be count-based (number of data items) or time-based (data items within an interval). The window slide defines the intervals the operator starts a new window on the input stream. It, too, can be time- or count based or rely on a predicate \cite{Arasu:2006:CCQ:1146461.1146463, mayer_predictable_2015, Grossniklaus:2016:FDW:2933267.2933304}.
For a deeper discussion of windows and classification of windowing policies, we point to the literature~\cite{hummer_elastic_2013, doi:10.1002/spe.2194}.

\subsubsection{Infrastructure Model}
\label{sec:infrastructure}
The infrastructure on which an SP system is deployed drastically influences the system's performance and thus the need and capability to parallelize the processing. This applies in particular to the type of processing nodes and the memory architecture.

\paragraph{Type of Infrastructure} Infrastructure types for SP systems are single nodes, cluster, cloud or fog solutions. Single node solutions run on a single, multi-core machine with scalability limited by the machine size. An SP system can scale up only, i.e. add more threads, as long as the node size permits. Clusters provide a fixed set of processing nodes. The cluster size limits the SP systems scalability. A common optimization objective for single node and cluster solutions is high resource utilization. SP systems running in a cloud environment have less limited scalability. They face the trade-off between processing performance and cost. The communication between stream sources and the cloud can induce a significant latency which can be critical in low-latency SP applications. Finally, the a fog infrastructure is limited in scalability, too, but often provides low communication latency as the processing can be performed close the sources. Heterogeneous processing nodes might influence the processing speed of the SP application in all types of infrastructure.

Some solutions explicitly integrate specialized hardware, especially GPUs and FPGAs. Exploiting this hardware properly requires additional efforts in the system design. For instance, GPUs provide a high throughput for highly parallel problems, but incorporate a latency and bandwidth penalty for first transferring the input events from the host memory to the device memory. 
 
\paragraph{Memory Architecture} Generally, SP operators communicate asynchronously via message passing. Some systems support operations on shared memory for communication and state management, e.g., when multiple operators are placed on the same host. This reduces communication and state-migration efforts \cite{mayer_spectre} but has higher access synchronization efforts.

\subsubsection{Operator State Models}
\label{sec:operatorstate}

SP systems further differ in their operator state model, i.e. whether and how they support stateful processing and state management in general and per operator. Operator state models are \textit{stateless} or \textit{stateful}. It is common that SP systems comprise stateless \textit{and} stateful operators at the same time. 
The operator state model 
influences if, where in the operator graph and to what extend parallel processing is possible. Additionally, it influences the synchronization and access coordination overhead that parallel processing might induce. 
\paragraph{Stateless Operators.}
Stateless operators consider one single data item at a time. It does not store results or information from formerly processing data items; instead, it treats each single data item the same way, regardless of former processing. Examples for stateless operators are filters on temperature data or rescaling of frames in video streams. 
\paragraph{Stateful Operators.} A\textit{ stateful operator} stores received data items or intermediate results as state. It uses and updates this state when it processes subsequent data items. An example is a fire detection application that raises an alarm if the temperature is above 50 degrees Celsius and smoke has been detected in the same area \textit{before}. 

The state an operator manages can be limited in scope and lifetime. The scope is limited when an operator processes only the sub streams of a certain set of keys and thus only keeps the respective state for this key set.
Time is limited when state is kept for window-based sub streams and dropped once the window is processed. An example for window-limited state is to calculate the average temperature of the last week from daily temperature measurements. It is possible to combine both dimensions, e.g., to limit an operators  by a key and additionally limit the lifetime of the state with a window model. A detailed model of partitioned state in SP operators is available \cite{DeMatteis:2016:KCR:2851141.2851148}.

\paragraph{State Management}
\label{sec:statemanagement}
State management defines if an SP system externalizes state of operators. 

\emph{Externalized State.}
To externalize state, an SP system provides a way to access internal operator state, e.g. via an API or a shared data storage. For instance, in operators with key-partitioned state, internal operator state is externalized to a key-value store~\cite{CastroFernandez:2013:ISO:2463676.2465282}. 
Managed state access become a bottleneck. A discussion of this issue, including a mathematical modeling of the problem, is provided by Wu et al. in \cite{wu_parallelizing_2012}. A practical example of applying such a scheme is provided by Hochreiner et al. \cite{7820260} in their distributed SP system PESP. Finally, Danelutto et al.~\cite{doi:10.1177/1094342017694134} provide a systematic classification of state access patterns in SP systems.

\emph{Internal (Hidden) State.}
In most SP systems, the internal operator state stays hidden to the rest of the SP system (c.f. Section \ref{sec:categorizationChapter}). While this avoids access management,
it hampers parallelization, elasticity, load balancing and state migration.

\subsection{Operator Parallelization Methods}
\label{sec:parMethods}

In the following, we provide a detailed introduction of two parallelization concepts for SP operators: \textit{task parallelization} (Section~\ref{sec:taskParallelization}) and \textit{data parallelization }(Section~\ref{sec:DataParallelization}). 
\subsubsection{Task Parallelization}
\label{sec:taskParallelization}

\begin{figure}
    \centering
    \begin{minipage}{0.4\textwidth}
        \centering
	\includegraphics[width=0.95\textwidth]{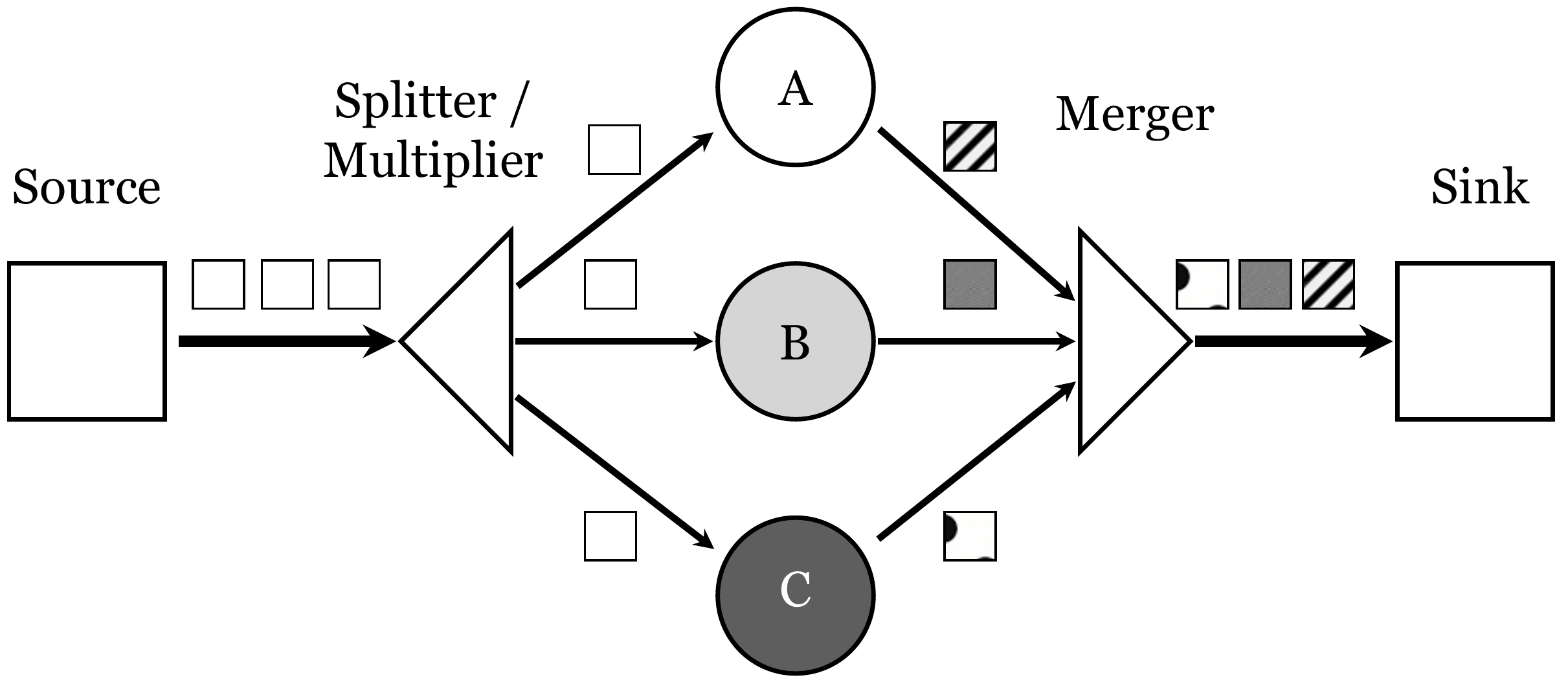}
	\caption{In task parallelization, an SP system runs  different operators in parallel. The splitter or multiplier distributes incoming events to all operators. The merger summarizes the results into one output stream and forwards it to the sink.}
	\label{fig:taskParallelization}
    \end{minipage}
	\ \ \ \ 
    \begin{minipage}{0.5\textwidth}
        \centering
	\includegraphics[width=0.95\textwidth]{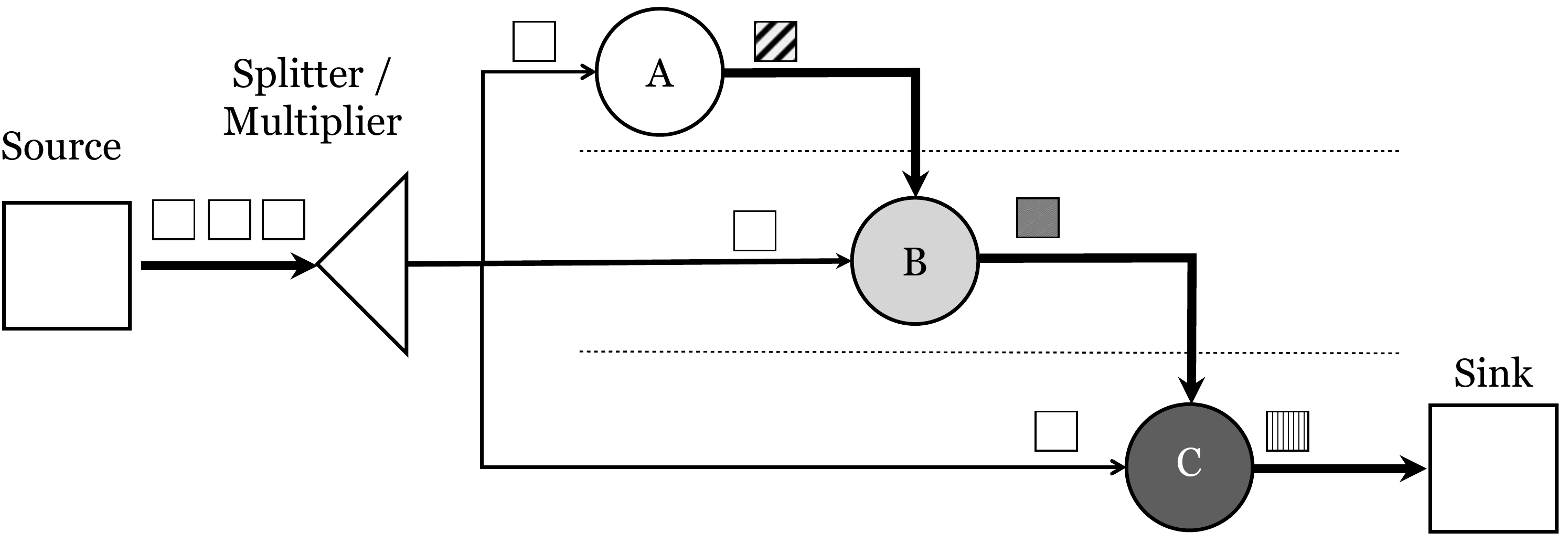}
	\caption{In pipelining, each operator receives processed the output events its preceding operator. Sometimes, an operator additionally received a copy of the original input stream.}
	\label{fig:pipelining}
    \end{minipage}
\end{figure}

In task parallelization, the SP system runs multiple operations on the same input stream in parallel. As an example, see Figure \ref{fig:taskParallelization}: A multiplier replicates the input data items and sends the replicas to operators A, B and C-The merger unifies the resulting data streams into one output stream.
An example for task parallelism is the encoding of a video stream in a live-streaming situation into different formats in parallel. Task parallelization enables pipelining where the output of one operator is the input of the next operator. In the context of SP systems, pipelining splits up bigger operators into consecutive sub-operators that can run in parallel (cf. Figure \ref{fig:pipelining} where an operator that detects a sequence ``A'', ``B'', and ``C' is split into three sub-operators). 
A requirement for task parallelization is that multiple operations (i.e. tasks) can run in parallel on the same input. 
Applicability thus depends on the concrete application.

\subsubsection{Data Parallelization}
\label{sec:DataParallelization}
\begin{figure}
	\centering
	\includegraphics[width=0.5\textwidth]{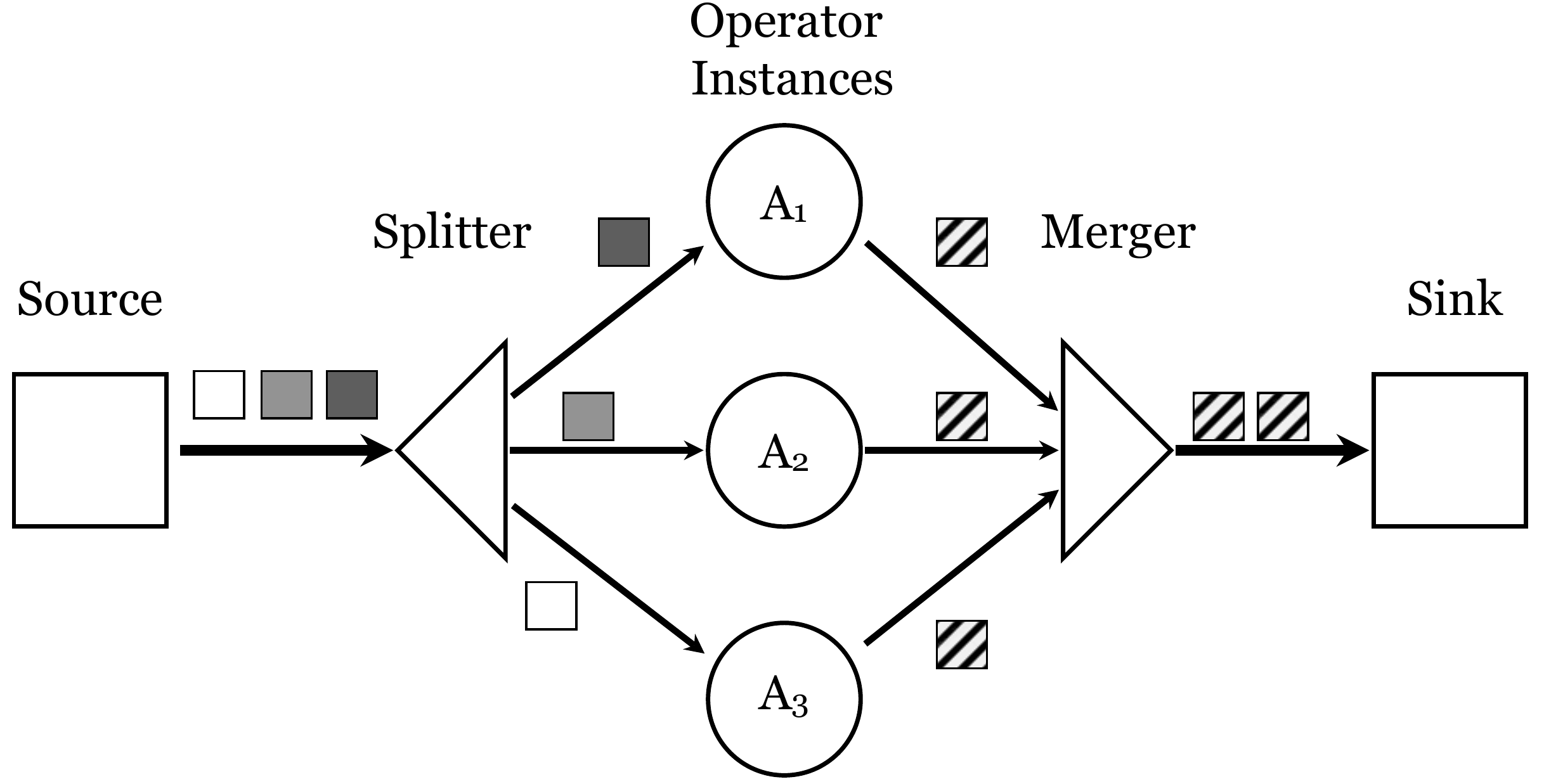}
	\caption{In data parallelization, a splitter divides the input stream and distributes the data items among instances, i.e., identical copies, of an operator. The instances process in parallel their assigned part of the input stream, i.e., they all perform the same operation on different parts of the data. A merger receives results from each instance and bundles them into one output stream.}	
	\label{fig:dataParallelization}
\end{figure}

The second parallelization method is data parallelization. It executes multiple instances of an operator, i.e., identical operator copies, in parallel on different parts of the input data. The number of instances is the \textit{parallelization degree} of the operator. To enable data parallelization, the input streams needs to be partitionable.  A \textit{splitter} component splits the input stream into sub-streams. The splitter can be a process on its own or integrated into operator instances, depending on the splitting strategy. 
 A \emph{merger} aggregates the instances output again into a single stream and, if necessary, ensures in-order delivery to the downstream operators.  Figure  \ref{fig:dataParallelization} shows the basic architecture of a data-parallel SP operator.

Stateful operators require special attention in data parallel SP systems: The input stream should be distributed among the operator instances such that each instance can keep an individual state. This avoids interference in between the different operator instances. 

In the following, we describe the three types of splitting strategies common in data parallelization:
\paragraph{Shuffle Grouping}
\label{sec:statelesssplitting}

With \textit{shuffle grouping}, the splitter shuffles data items across the operator instances. This is applicable in particular for stateless SP operators (cf. Section \ref{sec:operatorstate}), that process data items independent from each other. The splitter itself can be stateless, too, making an independent assignment to an instance for each single data item. A common example is Round-Robin splitting that assigns input data items to the operator instances in a Round-Robin fashion. 

If the SP operator implements an associative function, shuffle grouping can also be applied to operators that have key-partitioned state \cite{7113279}. Each operator instance keeps its own state for each key it has received. A combiner periodically combines the states of each key to the complete state. An example is a word counting application where each instance counts appearances per word and the combiner then calculates the word-wise total. Depending on the number of keys and instances, the combine stage can become very expensive. 

A stateful operator can implement shuffle grouping if all instances can access the complete operator state. The order of state access shall unrestricted to avoid sequential processing~\cite{doi:10.1177/1094342017694134}.

\paragraph{Key-based Splitting}
\label{sec:keybasedSplitting}

\begin{figure}
	\centering
	\begin{minipage}{0.45\textwidth}
		\centering
		\includegraphics[width=0.95\textwidth]{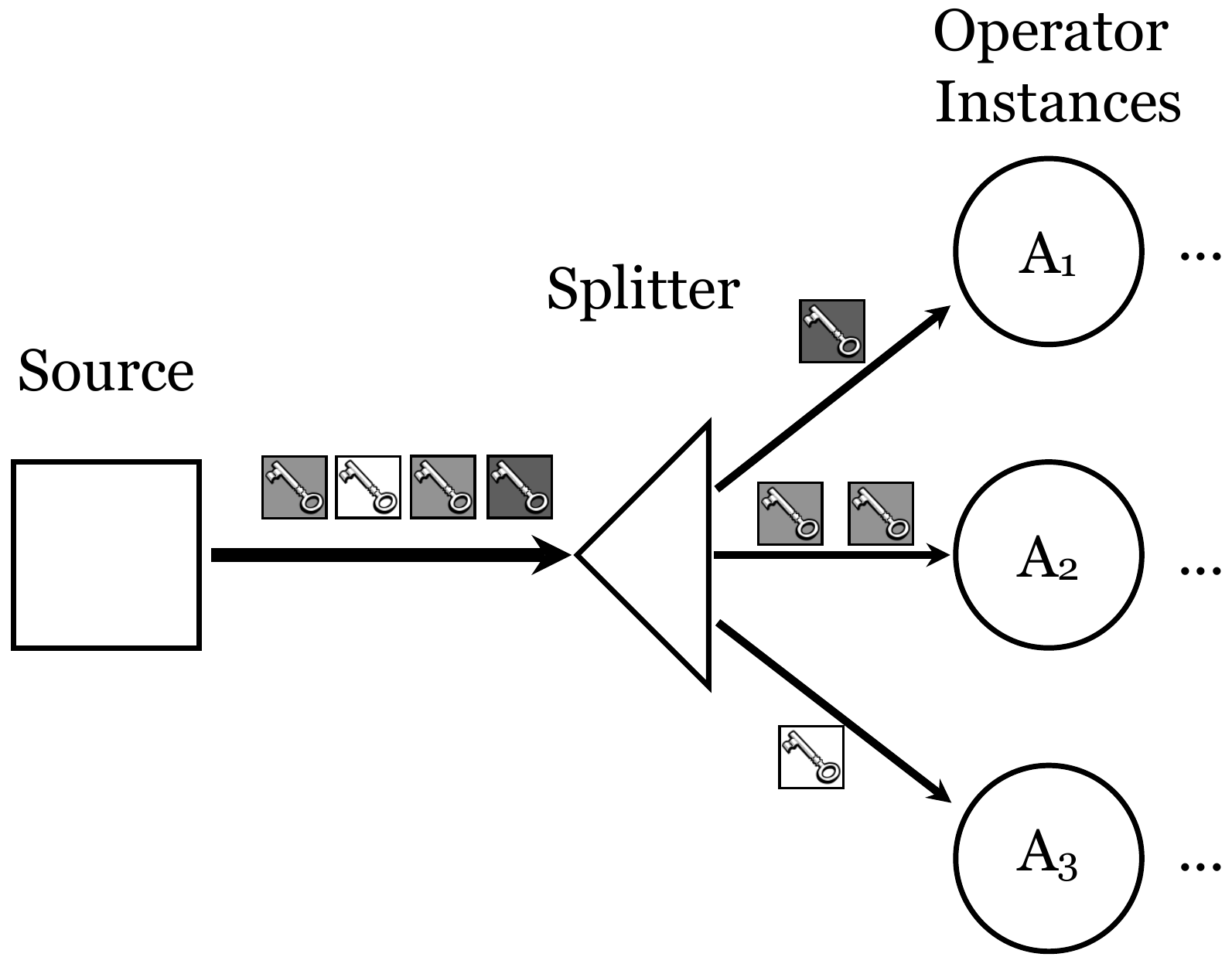}
		\caption{In key-based splitting, the splitter divides the input stream based on keys. These keys are attributes of the events. Each operator instance is responsible for a sub-range of the total key set.}
		\label{fig:keySplit}
	\end{minipage}
	\ \ \ \ 
	\begin{minipage}{0.485\textwidth}
		\centering
		\includegraphics[width=0.95\textwidth]{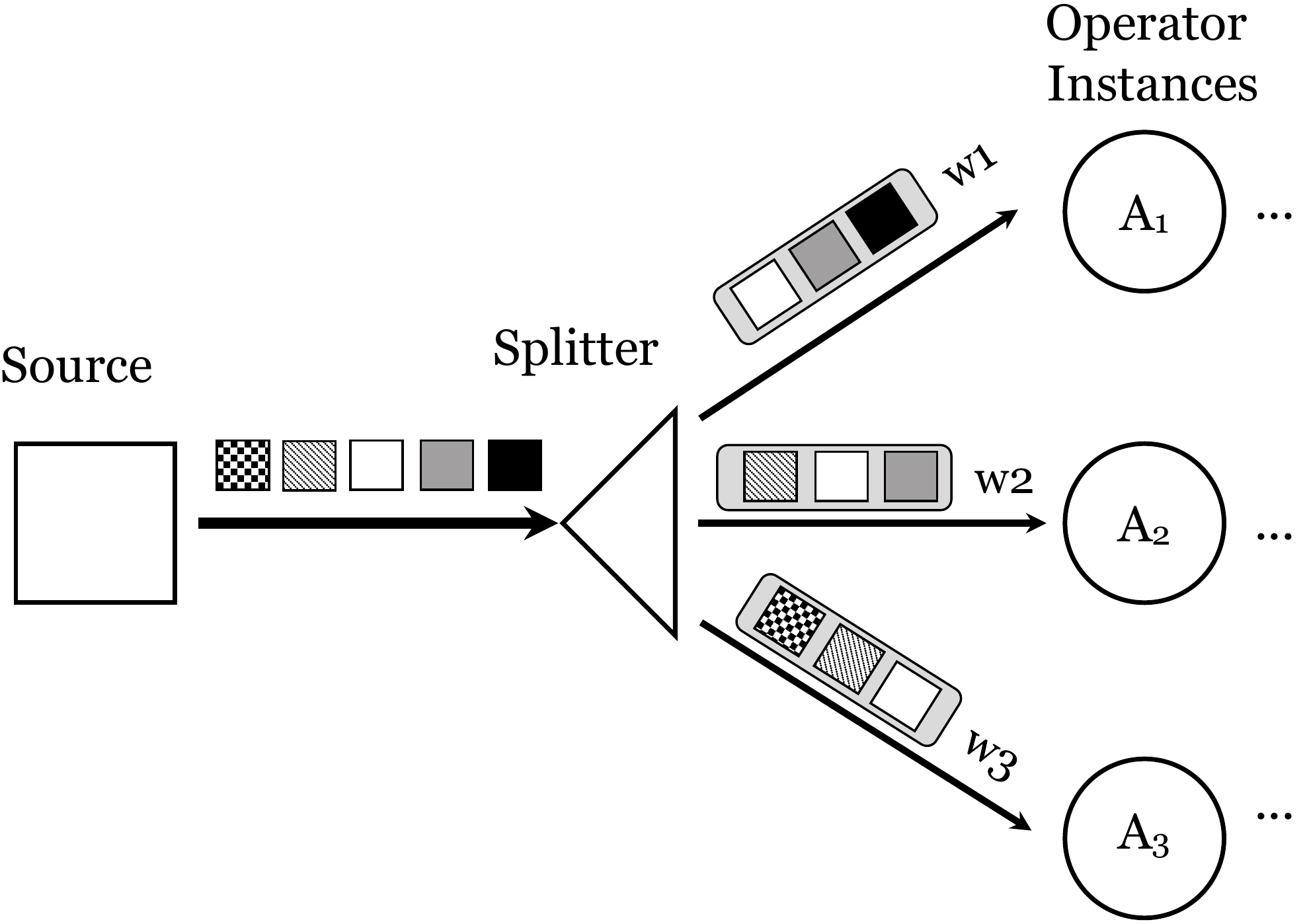}
		\caption{In window-based splitting, the splitter divides the input stream into windows of subsequent events. Each operator instance processes a sub set of all windows.}
		\label{fig:windowSplit}
	\end{minipage}
\end{figure}

Key-based splitting is applicable if operators are stateless or manage state per individual key. 
Balkesen et al. \cite{balkesen2013adaptive} therefore name it content sensitive splitting as opposed to content insensitive splitting like window-based or shuffle.  In the following, we simply denote the value of a key parameter itself as ``key''. Different ranges of keys are assigned to different operator instances, such that each operator instance keeps the state of a distinct, non-overlapping key range. A schematic of key-based splitting is depicted in Fig. \ref{fig:keySplit}. The different gray-scales of the events visualize their key-range, e.g., a specific stock symbol. The splitter forwards data items with the same key-range to the same operator instance.

\paragraph{Window-based Splitting}
\label{sec:winsplit}

In \textit{window-based splitting}, the splitter partitions the input event stream into subsequences, i.e windows of data items. It then assigns the windows to the instances of the operator (cf. Fig. \ref{fig:windowSplit}).
Depending on the window policy, Li et al. \cite{Li:2005:SET:1066157.1066193} differentiate between two types of \emph{context}---backward context and forward context---needed in order to determine which windows a given data item belongs to. Backward context of an data item $e$ may contain any information about previous data that arrive at the operator, whereas forward context refers to information from subsequent data in the input stream after $e$. In backward context approaches, when an data item $e$ is processed in the splitter, a new window can immediately be opened and scheduled to an operator instance, if applicable. However, this is not feasible if the window policy requires forward context.

\begin{figure}
	\centering
	\includegraphics[width=0.65\textwidth]{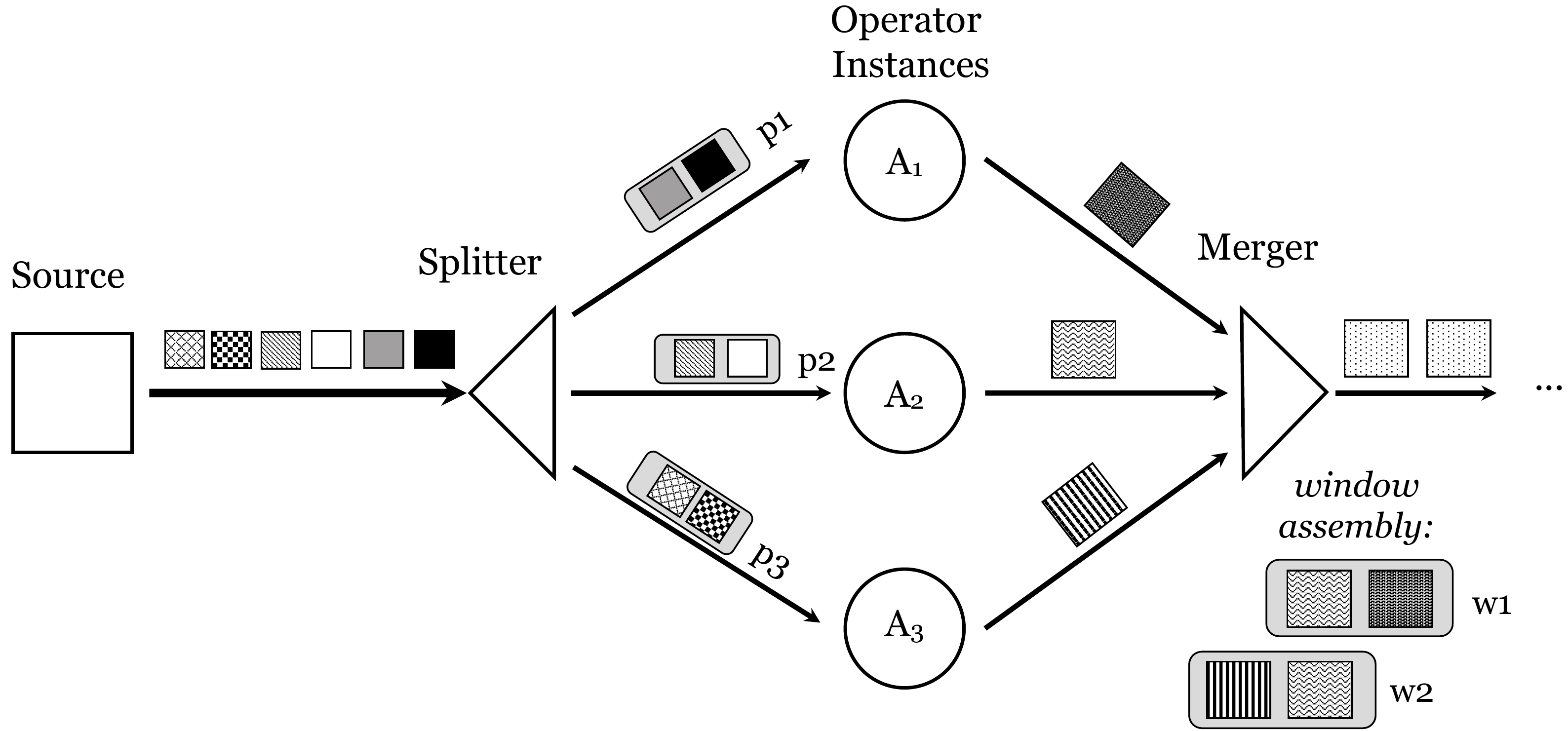}
	\caption{In pane-based splitting, the splitter divides the input stream into non-overlapping sub-sequences of events, called panes. Each pane belongs to one or more windows. The merger assembles the results according to the windows they belong to. }
	\label{fig:paneSplit}
\end{figure}

\paragraph{Pane-based Splitting}
\label{sec:panesplit}
We have discussed in Section \ref{sec:winsplit} that window-based splitting can increase communication overhead for overlapping windows. Furthermore, processing each window from scratch is often not necessary and parts of the computation results in overlapping windows could be shared by all windows. \textit{Pane-based splitting} has been proposed by Balkesen et al. \cite{balkesen_scalable_2011} and Li et. al \cite{li_no_2005}. Pane-based splitting partitions the input stream into non-overlapping sub-sequences, called panes; each pane belongs to one or more windows (cf. Fig. \ref{fig:paneSplit}). The panes are processed independently and in parallel by operator instances. 
After the processing, the merger assembles the received results from the panes according to the window policy of the operator. For instance, when the max temperature value in a window with a scope of 1 minute and a shift of 10 seconds shall be computed, the input stream can be split into panes of 10 seconds each. The operator instances compute the max value in each pane in parallel and send the results to the merger. The merger computes the max value of a specific window by determining the max value from all 6 panes that belong to that window. 

\subsubsection{Limitations of Operator Parallelization Methods}
In the following, we briefly discuss each of the presented parallelization methods. We aim to give a better understanding of the opportunities and drawbacks of each method. 
\paragraph{Limitations of Task Parallelization}
Whilst being a well-established parallelization method, task parallelization, including pipelining, has three major drawbacks: First, it can increase network traffic when each operator has to receive the complete input stream. Second, there is the risk of load imbalance between the operators if they differ in processing speed. 
Third, the scalability of task parallelization is limited: A given operator can only be divided into a limited set of sub-operators until an atomic operation is reached or, which is more likely, the cost of distributing the processing outweighs the gains drawn from further parallelization.

\paragraph{Limitations of Data Parallelization}
The major challenge in data parallelization is to achieve well-balanced workload and  in-order delivery of data items to downstream operators. 
Key-based data parallelization has three limitations: Expressiveness, scalability and load balancing. Expressiveness and scalability are limited by the the number of distinct partitions, i.e. distinct key-values. For instance, when checking the stock market for stock patterns, the key-based parallelism is constrained to the number of distinct stock symbols. Further, key-based splitting is applicable only if the data items provide the respective key-values. 
Explicit load balancing between the operator instances is required when the keys are distributed unevenly across the input data. Situations can occur where one instance experiences a high arrival rate while other instances run idle  \cite{rivetti2015efficient}.

Window-based splitting might increase communication overhead: When different overlapping windows are assigned to different operator instances, the data items from the overlap of those windows have to be replicated to all corresponding operator instances. To reduce this overhead, \emph{window batching} assigns multiple subsequent overlapping windows to the same operator instance \cite{balkesen_scalable_2011, matteis_parallel_2017, Mayer:2017:MCO:3093742.3093914}. 
The strength of window-based splitting is more flexibility in the input data structure, e.g. does not require keys for splitting. 

Pane-based splitting requires the operator function to be dividable into two stages: One stage where the single panes are processed, and one stage where the windows' results are assembled from several panes' results.  
However, it does not require associative operations  as opposed to key-based splitting: the system still processes the data items in sorted groups in the panes, preserving the ordering of the items. 

\subsection{Operator Elasticity Methods}
\label{sec:elMethods}
The employed parallelism of an SP System shall ensure the primary QoS goals (c.f. Section \ref{sec:generalModel}). As SP Systems are usually long-running, circumstances change: Workload increases or decreases, the hosting infrastructure might have to be shared with other applications and also the content-related properties, e.g. the distribution of keys, can change. For instance, in traffic monitoring, rush hours impose higher workload than low traffic at night. 
To meet its QoS goals, an SP system has to provide parallelization accordingly, e.g. high parallelism for high workload. Still, a high degree of parallelism requires resources and it is wasteful to \textit{always} provide those resources needed to handle potential workload spikes. Latest with the surge of cloud computing and its pay-as-you-go business model, it becomes attractive to keep the reserved resources as minimal as possible. 
Thus, running SP systems comes with the continuous task to define the required degree of parallelism to meet its QoS goals while being resource minimal. Therefore, research developed methods to adapt the \emph{degree of parallelism} according to the current circumstances while minimizing resource costs of the SP system. A precise adaptation thereby is the optimal balance between meeting the QoS goals, and minimize the required resources, i.e. the cost. These SP systems, that can adapt the degree of parallelism at runtime, are called \textit{elastic SP systems}. 

Elasticity, i.e. adapting parallelism, changes resource requirements of the SP system at runtime,  e.g. new processing nodes are required for parallelization increase and instances need to shut down due at a decrease. Therefore, adapting parallelism requires to define how resources can be --- fast and efficient --- acquired and released, e.g. with idle nodes or anticipated future node requirement and according placement, scheduling and migration strategies. While this survey focuses on parallelization strategies, we mention the former in the discussion in Section \ref{sec:elasticity} where applicable and point to related work (Section~\ref{sec:related})s for more detailed approaches.
Finally, the adaptation overhead needs to be balanced with the adaptations benefits. 

In the following, we discuss the properties of elasticity approaches that we use to categorize elastic SP systems in Section \ref{sec:elasticity}.

\subsubsection{Input Data} Elasticity methods base on data from \textit{system information} and \textit{workload information}.
System information comprises CPU utilization, throughput, latencies for queuing and processing as well as memory consumption. Workload information is the number of data items in the input streams and their data type.
What information is chosen depends mainly on information availability, required precision level and optimization objective.
The information can be per individual operator instance, processing node, for a sub-graph or the complete application. Often, the elasticity component uses aggregated data, e.g. averages, predicted trends or a learned distribution.

\subsubsection{Timing}
The timing when an elasticity method adapts the SP system can be \textit{reactive} or \textit{proactive}.
Reactive approaches adapt when the system detects that its QoS goal is violated. In proactive approaches, the system anticipates a violation and adapts \textit{before} it violation occurs, e.g. with prediction models. Reactive approaches adapt the system according to measured data and can thus adapt the parallelization degree to meet the current workload accurately. They are usually simpler because they lack a prediction model. However, due to the delay until the adaptation is fully effective, they might temporarily lead to over-utilization of operator instances, data loss, or SLA violations. Proactive approaches model a future system state and can thus prevent QoS violations. However, the quality of the prediction strongly influences the approaches precision. We also find hybrid approaches that combine reactive and proactive elasticity.

\subsubsection{Objective} Each elasticity method has an optimization objective limited by side conditions. The objective aligns with the system's QoS goals, e.g. maximize throughput or minimize latency. 
Approaches that target high utilization are contrary to approaches that maximize throughput or minimize latency, because high utilization hampers these QoS goals.  
The side conditions are usually defined by a cost model that describes the SP system's cost, e.g. in terms of used resources.
Additionally, adaptation costs can be considered, e.g. if the system halts processing during adaptation which leads to latency spikes. In case of adaptation costs, 
system stability becomes an important side condition. Some approaches add balanced load, node utilization and fault tolerance to their objective. 
If noteworthy, we shortly present the cost model when we describe the approaches.

\subsubsection{Guarantees} Most elastic approaches keep their QoS goals in a best effort manner, i.e. on average over a wider amount of time. However, some provide real-time guarantees. This includes probabilistic soft-guarantees, e.g. that the application keeps in 90\% of the time its latency limit. 

\subsubsection{Methodology}
The methodology defines how an elastic approach comes to its adaptation decisions. Methodology types are \textit{threshold policy driven}, \textit{model driven }and  \textit{learning based}. Threshold-policy approaches directly compare the input data against a set threshold.  
An expressive threshold ensures that elasticity is triggered neither too early nor too late. While thresholds facilitate decision making, finding expressive thresholds, e.g. with profiling, is challenging.
Model-driven approaches use the input data to calculate with a mathematical model the new degree. These approaches face the challenge  of finding model that represents the system's real behavior.
Learned base approaches \textit{learn} a model from measured or profiled data that. These models are usually refined at runtime which improves precision but requires a critical amount of data for learning.

\subsubsection{Centralized and Distributed}
Elasticity approaches differ in \textit{where} the adaptations are controlled: Many approaches are centralized, i.e. one central component manages the parallelism for the complete operator graph. A centralized component has the global view and can thus thrive towards a global optimum. However, the central component can become a bottleneck and introduce communication overhead. Thus, some solutions implement a distributed approach.

\subsubsection{State migration}
Changing the parallelization degree of a stateful operator might require state migration.
Consider, for example, an application with key-based states where each operator instance is assigned a key range and processes those data items whose key falls into the instance's range. If the parallelization degree, hence the number of instances, changes, so has to the distribution of the key ranges. 
The respective state of the re-distributed keys needs to be re-distributed accordingly.
To manage these state migrations, most elasticity approaches provide a state-migration protocol, e.g., \cite{DeMatteis:2016:KCR:2851141.2851148, DEMATTEIS2017302}. To realize the migration and avoid inconsistencies, in most SP systems, the processing stops completely or partially during the state migration. Stopping the processing however adds a waiting latency. 
A common optimization dimension for elastic SP systems is therefore to minimize the number of migrations or the related downtimes \cite{heinze2014latency, cardellini2018decentralized}. 
One solution to minimize the number of state-migrations for key-based states is \emph{consistent hashing}. 
A consistent hashing function ensures balanced load and minimal state migrations when it (re-) distributes key-ranges upon adding and removing nodes \cite{balkesen2013adaptive, Karger:1997:CHR:258533.258660}. 
Additionally, Gedik \cite{gedik_partitioning_2014} proposes partitioning functions that even outperform consistent hashing. 

To enable state migration, many key-based SP systems use the Flux-operator in their splitter \cite{shah_flux:_2003}. It supports re-partitioning of key-ranges, state migration and load balancing. 
A buffer absorbs short-term imbalances. For long-term imbalances the buffer cannot absorb, Flux re-distributes the key-ranges to remove the imbalances and consistently migrates the affected state. 
To keep consistency, the processing at the instances involved in the migration stops and their input is buffered. 
Due to this buffer, those instances not involved in migration can continue processing. 

Shukla and Simman recently proposed two solutions for a reliable and fast \textit{execution} of state migrations \cite{shukla2018towards} for the SP system Storm \cite{ApacheStorm}. Storm natively stops operators and drops input-queues before migration. However, this induces the need for frequent checkpointing and data-item replay to ensure consistent stream processing. The author's first solution stops sources from emitting further data items, processes all input queues before stopping the tasks, checkpoints the state and re-uses it after scaling is finished. This solution avoids re-play of data items and requests checkpointing at migration time only.
Their second solution further reduces latencies with faster checkpointing. Additionally, input and ouput queues are stored together with the checkpoints to be resumed by the new instances responsible.

\subsubsection{Evaluation}
Some elasticity approaches evaluate the effect of an adaptation. They continuously improve their decision making process. Examples are a black-list for non-effective adaptations or model updates of learning agents. Again, we highlight those approaches in our discussion in Section \ref{sec:elasticity} where applicable.

\section{Categorization of Parallelization and Elasticity Approaches}
\label{sec:categorizationChapter}
This section discusses approaches for parallel and elastic SP. Further, two tables give an overview and how the approaches are classified according to the properties introduced in Section \ref{sec:parallelizationClasses} and Section \ref{sec:elMethods}, respectively. 

\subsection{Parallelization}
\label{sec:parallelization}

\begin{table}[]
	\footnotesize
			\begin{tabular}{@{}|l||p{0.8cm}|p{1cm}||p{1cm}||p{2.8cm}|@{\hspace{0.08cm}}p{0.22cm}||@{\hspace{0.08cm}} p{0.22cm}|@{\hspace{0.08cm}}p{0.22cm}|@{\hspace{0.08cm}}p{0.22cm}|@{\hspace{0.08cm}}p{0.22cm}|@{\hspace{0.08cm}}p{0.22cm}|@{}}

	\hline 
& \multicolumn{2}{c||}{Processing} &  & &  &\multicolumn{5}{c|}{Parallelization} \\ 
	 Ref, Name& System Type & Prog. Model & External State & Infrastructure & \noindent SM & TP & SG & KS & WS & PS \\ 
\hline 
\hline
	Esper \cite{Esper} & CEP & Dec &   & Cluster &x &  & x & x & x &  \\ 
\hline 
	Storm \cite{ApacheStorm} & GP & Imp &  & Cluster, Cloud, Fog &  & x & x &x& x & \\ 
\hline 
	Heron \cite{kulkarni2015twitter} & GP & Imp &   & Cluster, Cloud, Fog  &  &x& x & x & x &   \\ 
\hline 
	Spark \cite{Zaharia:2013:DSF:2517349.2522737} & GP & Both & &  Cluster, Cloud, Fog&  &x& x & x & x &     \\ 
\hline 
	Flink \cite{Carbone2015ApacheFS} & Both & Imp &   & Cluster, Cloud, Fog &  & x & x & x& x & \\ 
\hline
\hline
Thies, Gordon  \cite{thies_streamit:_2002,Gordon:2002:SCC:605397.605428, Gordon:2006:ECT:1168857.1168877}	& GP & Imp & Stateless & Grid & x & x & x & & &\\ 
\hline 
Cherniack  \cite{cherniack_scalable_2003}	 & GP & Dec & & Cluster &  & x & x & x &  & x \\ 
	\hline 
 Brenna \cite{brenna_distributed_2009}	& CEP & Dec &   & Cluster  &  & x  & & x &  &  \\ 
\hline
  Khandekar \cite{Khandekar:2009:COS:1656980.1657002}& GP & Imp  & & Cluster (he) &  & x & & & &  \\ 
\hline
  Woods \cite{woods_complex_2010}& CEP &Dec & External & FPGA &  & & & x & &\\ 
\hline   
  Neumeyer \cite{neumeyer_s4:_2010}& GP & Imp &  & Cluster, Cloud &   &  & & x & & \\
\hline
	Andrade  \cite{andrade_processing_2011}& GP & Imp & & Cluster &   &  & &x &x &\\ 
	\hline 
	Balkesen \cite{balkesen_scalable_2011} & GP  & Imp &  & Cluster  &   &  & & & x & x\\ 
	\hline 
  Zeitler \cite{zeitler_massive_2011}& GP & Imp &  & Cluster &   &  & & x & & \\ 
	\hline 
 Schneider	\cite{schneider_auto-parallelizing_2012} & GP & Dec &  & Cluster &    & &x & x & & \\ 
	\hline 
  Gulisano	\cite{gulisano_streamcloud:_2012}& GP & Dec &  & Cluster  &    & x & & x & & \\ 
	\hline 
  Wu \cite{wu_parallelizing_2012}& GP & Dec & External & Cluster & x &  & x & x & & \\ 
	\hline 
  Hirzel \cite{hirzel_partition_2012}& GP &  Dec  & External  & Cluster &    & & & x & & \\ 
	\hline 
 	Cugola \cite{cugola_low_2012}& CEP  & Dec & External  & Single Machine & x & x & & x & & \\ 
	\hline  
 Fernandez	 \cite{CastroFernandez:2013:ISO:2463676.2465282}& GP & Imp & External  & Cloud &    &  & & x & & \\ 
	\hline 
  Balkesen \cite{balkesen_rip:_2013}& CEP & Dec  &  & Cluster  & x & x & & & x & \\ 
\hline  
  Balkesen \cite{balkesen2013adaptive}& GP  & Imp  &  & Cluster, Cloud &   &  & x & x & & x \\ 
\hline  
  Wang	\cite{wang_complex_2013}& CEP & Dec &  & Cluster  &    & x & x & &  & x \\ 
\hline  
Tang \cite{tang2013autopipelining} & GP & Imp & & Single Machine &  & x &  & & & \\
\hline
  Fernandez \cite{Fernandez:2014:MSE:2643634.2643640}& GP  & Imp  & External & Cloud  &   &  & x & x & &   \\ 
\hline  
   Lohrmann \cite{lohrmann_nephele_2014}& GP  & Imp &  & Cluster, Cloud &   &x  &  & x & &  \\ 
\hline  
  Zygouras	\cite{zygouras2015insights}& CEP & Dec  &  & Cloud &   & &   & x & &  \\ 
	\hline
 	Schneider \cite{schneider_safe_2015}& GP & Dec  &  & Cloud  &    &  & x & x & &  \\ 
	\hline
  Rivetti \cite{rivetti2015efficient}& GP & Imp  &  & Cluster &   &  &   & x & &  \\ 
	\hline
Mayer \cite{mayer_predictable_2015,Mayer:2017:MCO:3093742.3093914} & CEP  & Imp  &   & Cloud &    & & & & x &\\ 
\hline
 Wu  \cite{chronostream}& GP & Imp  &(External) & Cluster &   & & & x & &\\ 
\hline
 Nasir  \cite{7113279, 7498273}&  GP &  Imp &   & Cluster &   &  & x & x & & \\ 
\hline  
 	Saleh \cite{saleh_partitioning_2015}& CEP  & Dec &  & Cluster  &   & x &  & x & & \\ 
	\hline  
 Koliousis	\cite{koliousis_saber:_2016}& GP & Dec &  &  GPU &x & x & & & &x\\ 
\hline  
	Zacheilas  \cite{zacheilas_dynamic_2016}& CEP &  Dec  &  & Cloud &   &  &  & x & & \\ 
\hline
 Nakamura	 \cite{nakamura_design_2016}& GP & Imp  &  & Fog &  & x & & & & \\ 
\hline   
 Gedik \cite{GEDIK2016106} & GP & Dec &  & Single Machine &x & x &  & x & & \\ 
\hline 
 Mayer  \cite{mayer_graphcep:_2016}& GP  & Imp &    & Cloud &    & & & x & x &\\ 
\hline 
 Rivetti  \cite{Rivetti:2016:OSS:2988336.2988347}&  GP & Imp & Stateless & Cluster &   &  & x & & & \\ 
\hline    
 Schneider \cite{schneider_dynamic_2016}	& GP & Dec  & Stateless & Cluster (he)&   &  & x & x & & \\ 
\hline   
 Katsipoulakis  \cite{Katsipoulakis:2017:HVS:3137628.3137639}& GP & Both &   & Single Machine, Cluster &  (x) &  & x &x & &\\ 
\hline  
 Mayer \cite{mayer_spectre}& CEP & Imp &  External  & Single Machine  & x &  &  &  & x  & \\ 
\hline 
Mencagli  \cite{mencagli2017harnessing}& GP &  Imp &  & Single Machine &x  &  &  & & x & \\ 
\hline       
 Mencagli \cite{7873332, MENCAGLI2018862} &GP  & Imp &  & Single Machine & x &  &  &  &  & x  \\ 
\hline 
\end{tabular} 

\caption{Categorization of SP operator parallelization literature. The "Processing" Columns show the type of SP system. ``State External" marks approaches that either externalize state or are limited to stateless operators. The Infrastructure section shows the target infrastructure and if the approach uses shared memory (SM). The last columns mark Task Parallelization (TP), Data parallelization with Shuffle Grouping (SG), splitting key-(KS), window-(WS) or pane-based (PS). 
	Open source frameworks are followed by research approaches in order of their date of publication.}
\label{tab:ParallelizationApproaches}
\end{table}

\label{sec:parallelizationCategorization}
In Table \ref{tab:ParallelizationApproaches}, we provide a systematic categorization of the literature on operator parallelization. The first column contains the name of the first author and the reference number in the bibliography. In the ``processing" columns, we show the system type and the programming model. As the majority of approaches uses internal state mangement, we highlight in the \textit{External State} column, if a system exposes the state of operators to other parts of the system. We further show the used infrastructure and the memory architecture. For the memory architecture, the default value is \textit{shared nothing}. If shared memory is assumed, we mark this with a tick in the column \textit{SM}. In the parallelization columns, we show the approaches type of parallelization. \textit{TP} is task parallelization, \textit{SG} for shuffle grouping, \textit{KS} for key-based splitting, \textit{WS} for window-based splitting, and \textit{PS} for pane-base splitting. 
All entries are sorted by the year of publication.

We can see from the table that the amount of work on operator parallelization started to boost around the year 2010, when cloud computing started its surge. Most research targets general SP. The popularity of data parallelization in research papers is very high with a clear dominance of key-based splitting.

In the following, we discuss approaches for SP parallelization in detail. We build three groups: Open source frameworks, parallelization approaches for CEP, and parallelization approaches for GP.  Within each group, we highlight the distinctive points in the respective systems or approaches, and provide a finer-grained grouping based on commonalities of the different approaches, if applicable.

\subsubsection{Open source frameworks}
\label{sec:frameworks}
Many parallelization and elasticity approaches we discuss in this article base their research on an open source SP framework. In this section, we present the most common ones. We explicitly focus on how they enable parallel SP. For each framework, we thus briefly discuss how they provide parallelization options for SP application developers. Only Spark inherently supports elasticity. The other frameworks have interfaces to control parallelism at runtime. For a more detailed discussion of these frameworks with a higher focus on their internal architecture we point to Assuncao et al. \cite{de2017resource}.
\textbf{Esper} \cite{Esper}:
Esper is a declarative SP system for CEP. Developers declaratively define topologies with EPL, a language similar to SQL. The open source version supports scale up and down on a multi-core node with four options for multi threading. 
Esper does not guarantee in-order processing when multi-threading. The input stream can be split with windows or context. Context can be built from keys, hash values, categories or time e.g. detect events between 9:00 AM and 5:00 PM. 

\textbf{Heron} \cite{kulkarni2015twitter}: 
Heron is a GP SP framework that supports key-based and window-based splitting. It has predefined splitting strategies an additional API for custom strategies.  
Heron provides multiple APIs to imperatively build SP applications. It is thereby API compatible with Apache Storm. 
Task parallel processing and pipelining can be incorporated by defining the topology accordingly.

\textbf{Storm} \cite{ApacheStorm}:
In terms of parallelization, Storm provides the same options as Heron does. Also input tuples and operator types are the same. Yet alone the API to define topology differs as Heron provides more API-options (e.g. the Streamlet API) than Storm does. Their major differences lie in architectural details that are out of the scope of this article. We point to Kularkni et al\cite{kulkarni2015twitter}.

\textbf{Spark Streaming and Structured Streaming by Spark}\cite{Zaharia:2013:DSF:2517349.2522737}: As Spark applications originally processed batches, the Spark Streaming extensions process streamed data in micro batches. \textit{Structured Streaming} interprets data streams as an unbounded table where each new data item extends the table.
Operators are queries on this table. A specific schema for input data is required. Data can be split using properties (i.e. keys) and aggregated as windows. Operators are defined declaratively. The parallelism degree can be set explicitly in the topology or via a default value in the configuration files. Opposed to the other frameworks, Spark has a Dynamic Resource Allocation module for elasticity. According to the documentation, it releases unused resources in low workload times and requests them again in peak workload times. 

\textbf{Flink}\cite{Carbone2015ApacheFS}: Flink supports key-based and window-based splitting. The parallelization degree for data parallelism can be set in the topology implementation or via an interface. As default, Flink places one instance per operator on each core. To the best of our knowledge, custom grouping mechanisms are not available. Besides a DataStream API,  Apache Flink provides two APIs - the Table API and SQL- API - to use relational-oriented query descriptions. For CEP-oriented queries, a CEP library is available. 

\subsubsection{Parallelization in CEP}
This section summarizes approaches for parallelization in CEP.

Brenna et al. \cite{brenna_distributed_2009} extend Cayuga~\cite{demers_cayuga:_2007}, a centralized, multi-query SP system for pattern matching. They enable pipelining and key-based data parallelization and distribute the processing of Cayuga. It resembles the pipelining approach of Balkesen et al. \cite{balkesen_rip:_2013} (cf. Section \ref{sec:taskParallelization}).   

Woods et al. \cite{woods_complex_2010} implement the CEP operators directly on FPGAs (Field Programmable Gate Arrays) for fast CEP. They focus on the so called ``network-memory-bottleneck''. This bottleneck occurs when a SP system writes the content of the event network packages to the main memory to be accessible by the CPU. The FPGA interprets the event network packages without writing them to main memory first. Only detected patterns are forwarded to the main memory. To exploit the parallelism of FPGAs, the authors introduce a query language that uses partition keys and predicates. The system keeps state per key.
As FPGA hardware limits the number of keys to be stored, the authors propose a time limit for the storage of keys. With this limit they discard the state of a key, if out of a fixed number of past events in the input stream, none has had the specific key. While the approach is less flexible and scalable due to its hardware-centric nature, it can be beneficial for highly latency sensitive applications.

Cugola and Margara \cite{cugola_low_2012} propose two algorithms for pattern matching in CEP operators. The first algorithm, ``automata-based incremental processing (AIP)'', detects patterns with a non-deterministic finite automaton, while the second algorithm, ``column-based delayed processing (CDP)'', uses lazy pattern matching. Both algorithms enable pipelining parallelism in multi-stage pattern matching operators, and task parallelism when multiple queries evaluate the same input event stream. An outstanding merit in the work of Cugola and Margara is that they take into account a heterogeneous computing environment with CPUs and GPUs. 
Based on evaluations, the authors recommend to use the GPU if there are few, but complex operators in the system, and multi-core CPUs for a high number of less computational-intensive, operators.

Balkesen at al. \cite{balkesen_rip:_2013} propose the window-based data parallelization approach ``Run-based Intra-operator Parallelization (RIP)'' for CEP. They implement the operator query as finite state machines (FSM) and start a new FSM instance for each incoming event. Additionally, each event is processed by the already running FSM instances. 
The authors propose to split the input event stream into overlapping batches of events based on the window policy of the operator. Each batch is assigned a thread that then runs those FSM instances started by the events in that batch. This approach requires bounded window sizes.

The system proposed by Mayer et al. \cite{mayer_predictable_2015} is a distributed, window-based data parallelization framework. Input event streams of an operator are split into windows that are scheduled to an elastic set of distributed operator instances, where each operator instance processes its assigned windows ``from scratch''. This leads to a very high expressiveness of the framework, as the authors show on operators defined in the Snoop \cite{chakravarthy_snoop:_1994} and CQL \cite{Arasu:2006:CCQ:1146461.1146463} query languages; basically, any window-based operator can be integrated into the framework. To expose the window policies of the operator to the splitter, a programming API has been integrated into the system, so that operators can be plugged into the system with minimal inference to the operator code.

Later work of Mayer et al.  \cite{Mayer:2017:MCO:3093742.3093914}  discusses the trade-off between communication overhead and load balancing when performing window batching. The authors propose a model-based batch scheduling controller that predicts the latency peak when assigning a window to an operator instance. Based on that, the controller assigns the windows to the operator instances in such a way that communication overhead is minimized while a latency bound in the operator instances is met.

Wang et al. present in \cite{wang_complex_2013} an approach that computes sequence pattern matches by splitting the event stream into uniformly sized batches, which resemble panes. Operator instances process those panes in parallel. The pattern matching is thus done per pane at first. Further, the relationships between events that build any sub-pattern of the complete pattern are stored.  A merge stage finds complete patterns by stitching together the sub-patterns reported from the operator instances. 

Hirzel \cite{hirzel_partition_2012} proposes a pattern syntax and a translation scheme for IBM's System S that supports pattern matching. An algorithm translates the match-expressions (the pattern to be found) into nondeterministic finite automata (NFAs) at compile time and generates C++ code for it. A ``partition map'' stores each NFA-state together with the aggregated results computed so far for the state. A ``partition-by'' attribute in the match-expression selects all states from the map that are relevant when a new event arrives. Parallelization is achieved by splitting the input stream on the same keys (partition-by attributes) that are used to select the partition maps. 

Zygouras et al. \cite{zygouras2015insights} implemented a highly distributed and parallel Big Data processing system that includes stream and batch processing capabilities. Special about this approach is that it combines the frameworks Esper \cite{Esper} and Storm \cite{ApacheStorm}. 

Saleh et al. \cite{saleh_partitioning_2015} consider stream and operator splitting of CEP applications.  As a basis, the authors use PipeFlow, their distributed CEP system that provides features for data parallelization~\cite{saleh_pipeflow_2015}. The input stream can be split based on keys. However, task parallelization with operator splitting is the focus of their work. 
The authors partition each CEP operator into sub-operators by rewriting the query according to rewriting rules. The rewritten query is then easier to split. 
To decide about the size of partitions, the system statically assigns costs for latency and memory consumption for each possible partition. A greedy algorithm then finds the cost-optimal set of operator graph partitions for the available set of computing nodes. 

Mayer et al. \cite{mayer_graphcep:_2016} developed GraphCEP, an SP system that allows for combining stream analytics with graph processing. Event streams are partitioned by a key and fed into operator instances. Each operator instance has access to a graph processing engine \cite{8263157} to perform heavy-weight parallel computations on a large, shared graph. Further, the merger allows for stateful computations that combine results from multiple operator instances.

In a window-based splitting approach, an event may belong to multiple windows due to a window overlap. \textit{Consumption policies} can define, that if an event contributes to one detected pattern, it cannot be part of another pattern detection, possibly in another window \cite{754955}. Hence, the processing of overlapping windows becomes interdependent. Some SP pattern definition languages that enable these policies are Snoop \cite{chakravarthy_snoop:_1994}, Amit \cite{Adi:2004:ASM:988145.988150} and TESLA \cite{cugola_tesla:_2010}. The SPECTRE system by Mayer et al. \cite{mayer_spectre} is the first work that addresses this interdependency by means of speculative processing. In particular, SPECTRE creates multiple versions of dependent windows that assume different event consumptions and assigns the most probable versions to operator instances. SPECTRE is designed for multi-core shared memory environments. 

\subsubsection{Parallelization for General Stream Processing}
 The following paragraphs discuss parallelization in general SP systems. Due to the high amount of approaches, we further group them according to the parallelization type, i.e. task or data parallelization. A special case are key-based data parallelization techniques: This biggest group of approaches is further split up for approaches that provide advanced key-partitioning functions, approaches that apply key-based splitting and approaches that focus on key-based state management. Please notice that these groups overlap and we assigned approaches according to the approaches focus.

\paragraph{Approaches implementing Task Parallelization}

The StreamIt language by Thies et al. \cite{thies_streamit:_2002} is a high-level programming language that supports both task parallelization and pipelining. StreamIt provides a programming abstraction for managing the event streams and operators. The compiler of StreamIt is described in two further publications \cite{Gordon:2002:SCC:605397.605428, Gordon:2006:ECT:1168857.1168877}, leveraging and optimizing task, pipeline and data parallelism.

COLA by Khandekar et al. \cite{Khandekar:2009:COS:1656980.1657002} is a pipelining optimization algorithm for IBM's System S. It optimizes queries at compile time, before the deployment of the operator graph. It fuses operators from the logical plan of the query into coarser-grained operators. The fused operators process the input stream in a pipelined way. The fusion balances communication cost and CPU capacities of the heterogeneous processing nodes.

Lohrmann et al. \cite{lohrmann_nephele_2014} propose Nephele-Streaming that extends their batch processing framework Nephele  \cite{warneke2011Exploiting}. To enable SP on top of Nephele, Lohrmann et al. adopt a \emph{micro batching} approach \cite{Zaharia:2013:DSF:2517349.2522737}. It processes data items in a stream of small batches. By changing the granularity of the operator graph, Nephele-Streaming enables pipeline parallelism in the operators.

Nakamura et al. present in \cite{nakamura_design_2016} an SP middleware that is based on their earlier system called ``Information Flow of Things'' (IFoT) \cite{yasumoto2016survey}. They designed a layer fog-computing architecture that uses Raspberry Pi-nodes. On these nodes run so called neurons that perform real-time data stream analysis. 
The middleware divides each application into tasks according to recipes that describe how the input data should be processed within the application. Tasks might be shared by different applications. Each neuron executes a set of tasks and can exchange information with other nodes via the connected "neuron layer". 
The input and output event stream of the sensors and actuators is managed by the neurons using the publish/subscribe paradigm~\cite{Eugster:2003:MFP:857076.857078}.

Tang and Gedik \cite{tang2013autopipelining} introduce task parallelism and pipelining in an SP system. In their system, each threads processes a sequence of operators as a sequence of function calls. To enable pipelining, an additional thread inserted into this sequence to execute the downstream operators and frees the upstream thread to process the next tuple. 
Their approach provides an elasticity mechanism we describe in Section \ref{sec:elasticity}. It runs on a single, multi-core node.

\paragraph{Approaches implementing Shuffle Grouping}

Approaches that implement shuffle grouping focus on load balancing.
Schneider et al. \cite{schneider_dynamic_2016}  propose an approach that balances load for parallelized stateless operators where the operator instances may differ in throughput. Their System S extension monitors the TCP blocking rate per connection between the splitter and each operator instance. Minimizing the maximal blocking rate among all operator instances balances the load.

Rivetti et al. \cite{Rivetti:2016:OSS:2988336.2988347} propose an approach that tackles imbalanced processing latencies of instances. They focus on applications where the data item processing latency can depend on the content and propose ``Online Shuffle Grouping'', an algorithm for proactive online scheduling of input data based on an estimation of the data items processing time.

Another splitting variant mixing key-based splitting and shuffle grouping is known as ``partial key grouping'', proposed by Nasir et al. \cite{7113279}. The shuffling phase is restricted to two deterministic options according to two different hash functions (i.e., key-based splitting with two hash functions instead of one). A number of parallel splitters can pick dynamically to which of the two operator instances to send an event, focusing on the one with less load. This optimization technique is known as ``the power of two choices'' and provides significant improvement in load balancing \cite{Azar:1999:BA:330358.330366}. Later, Nasir et al. extend their approach to allow for more than two choices for ``hot'' keys that impose most of the workload \cite{7498273}. In all of these approaches, a combiner is needed to combine the shuffled state of each key. 

An early attempt of key-aware shuffle grouping has been proposed by Balkesen et al. \cite{balkesen2013adaptive}. When an SP operator has key-partitioned state, the input stream is split by a variant of consistent hashing (``frequency-aware hash-based partitioning''). This variant considers the key frequencies. In particular, the $k$ keys with the highest frequency are further split into sub-keys that are shuffled around the operator instances---the number of splits depends on the frequency. 

Recently, Katsipoulakis et al. \cite{Katsipoulakis:2017:HVS:3137628.3137639} proposed to consider aggregation cost in the combiner when deciding where to route which key. Based on a mathematical formulation of imbalance \emph{and} aggregation cost (where aggregation cost directly depends on the number of operator instances that receive and process events with the \textit{same} key), they propose several heuristic splitting methods that aim to minimize both imbalance and aggregation cost for a given operator and workload.

In a data-parallel SP operator, the splitter can become a bottleneck if the input event rates are very high. Zeitler and Risch \cite{zeitler_massive_2011} address this problem by proposing a two-stage splitting processing. In the first stage, they split the input stream into fixed-size batches. They then route these batches to a number of parallel splitter instances. Those parallel splitters then perform the actual key-based event routing to the operator instances. The authors provide a mathematical model to compute the optimal batch size for the first splitting stage as well as the optimal number of parallel splitters. The approach works both for stateless splitters as well as key-based splitters.

\paragraph{Key-partitioning Functions}
This section summarizes SP approaches apply key-based splitting and put a particular focus on the key-partitioning function.

Rivetti et al.  \cite{rivetti2015efficient} propose an algorithm that learns which keys are currently the most common ones. Based on that, they apply a greedy algorithm to compute a nearly load optimal distribution, balancing frequent and rare keys among the available nodes. The same problem is also tackled by Zacheilas et al. \cite{zacheilas_dynamic_2016}, who formulate it as a variation of the job shop scheduling problem and apply an extension of the ``Longest Processing Time'' algorithm, a greedy heuristic solution.

In \cite{cherniack_scalable_2003}, Cherniack et al. deploy the SP system Aurora~\cite{Abadi:2003:ANM:950481.950485} in a distributed setting. 
To enable data parallelism, they split the operator (``box splitting'') and add a splitter and merger instance. The splitter (called ``filter with a predicate'') routes tuples based on predicates, giving the user advanced options to specify the splitting key with user-defined predicates. The merger can implement, e.g., a union of the output tuples or a sorting algorithm. The authors provide a deeper discussion on challenges of splitting and merging data streams.

Gulisano et al. \cite{gulisano_streamcloud:_2012} propose ``StreamCloud", an elastic and scalable SP system. Their key-based data parallelization method minimizes distribution overhead. They propose partitioning functions for different stateful SP operators (equijoin, cartesian product, and aggregate operator). 

\paragraph{State Management for Key-based Splitting}
Fernandez et al. \cite{CastroFernandez:2013:ISO:2463676.2465282} propose SEEP, a key-based data-parallel SP system. In SEEP, operators expose their internal, key-partitioned state to the SP system through a set of state management primitives. SEEP performs state management both for scale-out as well as recovery of operators. In their later work, Fernandez et al. \cite{Fernandez:2014:MSE:2643634.2643640} propose the abstraction ``stateful data-flow graphs'' (SDG) that separates data from mutable operator state. In their model, state elements can be partitioned by a key and dispatching is performed by hash- or range-partitioning on a key or by shuffle grouping. If a state element cannot be partitioned, it is replicated, and state updates are combined by a ``merge logic''. Further, Fernandez et al. provide a tool ``java2sdg'' \cite{7498352} that translates annotated Java programs to SDGs for execution in the SDG runtime system. 

Chronostream by Wu and Tan \cite{chronostream} splits the computational state (based on keys) into so-called slices. This state splitting supports horizontal and vertical scaling. The slices are distributed and replicated across different machines, leading to efficient load balancing and enabling fault tolerance. 

Wu et al. \cite{wu_parallelizing_2012} extend IBMs System S and enable it to handle shared state in data parallel processing. They implement a round-robin and a hash-based splitting routine. Special about their approach is a theoretical model. It analyzes if parallel processing improves the performance of an SP systems that employs shared state. The model predicts the waiting time and the time overhead that the shared state access induces.

\paragraph{Approaches implementing Key-based Splitting}
\label{sec:key-basedSystems}
Schneider et al. \cite{schneider_auto-parallelizing_2012} present a compiler and a runtime environment for their SP language SPL \cite{hirzel_spl:_2014} (Stream Processing Language) that enables data parallelization. Based on hints provided in SPL, the complier defines parallel regions, which are sequences of multiple operators and automatically replicates them. The system supports stateful and stateless operators and splits the input stream of a parallel region based on keys or, if the operator is stateless, with Round-Robin shuffle grouping. Fusion of operators further reduces network communication. In a later publication \cite{schneider_safe_2015}, the authors add support for operators that produce a dynamic number of output data items per input data item. 

Gedik et al. \cite{GEDIK2016106} propose an approach for ``pipelined fission'', a combination of pipelining and key-based data parallelization. They fuse groups of operators into so-called pipelines with a heuristic optimization algorithm. Different threads can then execute these pipelines independently to enable parallelism. Additionally, stateless or key-partitioning pipelines can be further parallelized by a split--process--merge architecture to exploit data parallelism. 

Neumeyer et al. \cite{neumeyer_s4:_2010} propose the highly scalable SP system S4. The system scales with Processing Engines (PE) where one PE is responsible for a specific key and processes all corresponding data items. A processing node (PN) manages multiple PEs, i.e., a subset of the key domain, and instantiates new PEs if a new key occurs that is not covered by an existing PE yet.

\paragraph{Approaches implementing Window-based Splitting}
Andrade et al. \cite{andrade_processing_2011}  present a low-latency SP implementation in IBM's System S. This implementation splits the input stream of an operator on two levels: On the first level key-based, 
on the second level, for each of the key ranges, window-based. Thus, it is for example possible to calculate an average over a time window, i.e. a window-based operation, for the stock value of different companies, i.e. on a key-based split input data stream.  The authors name this pattern the ``split/aggregate/join" pattern. 

 The work of Mencagli et al. \cite{mencagli2017harnessing} propose a novel execution model, ``agnostic worker'' for multi-core shared memory environments. An agnostic worker parallelizes operators with window policies that require forward context (cf. Section \ref{sec:parMethods}). The splitter does not only determine the window extents on the input stream, but also when a window computation is triggered (i.e., when the \texttt{query} function is called.). The actual computation of the window results, i.e., the execution of the query function, is then scheduled to one of the available operator instances.

\paragraph{Approaches implementing Pane-based Splitting}

Mencagli et al.~\cite{7873332,MENCAGLI2018862} split large panes into sub-panes with a proportional-integrative-derivative controller (PID) that automatically adjusts the splitting threshold. Their work  focuses on burstiness in event arrival rates as to avoid bottlenecks.

Balkesen et al. \cite{balkesen_scalable_2011} parallelize sliding window processing on data streams with pane-based splitting. A ring-based pane-partitioning splits the input stream and assigns panes to processing nodes that are logically ordered in a ring.
The ring structure eases ordered processing and reduces communication overhead. Their approach first assigns windows to the ring-ordered nodes round-robin. 
Each node then processes those panes that belong to its assigned windows.
A pane result calculated on one node can then be easily forwarded to the next node where the next window this pane belongs to assigned. A merge node aggregates the pane-results whereby it allows for a constrained degree of disorder to balance efficiency and output quality.

Koliousis et al. propose SABER \cite{koliousis_saber:_2016}, an SP engine that manages query processing on heterogeneous hardware with CPU and GPU cores. A splitter first splits the incoming event streams into batches of a fixed size and assigns them to a processing unit, i.e., a CPU core or a GPU processor. Each processing unit executes a ``query task'', i.e., a function that takes $n$ batches, one from each of the $n$ incoming event streams and calculates a function on those $n$ batches in parallel applying an $n$-ary operator function. SABER aims to keep the size of the batches independent from the window policy, i.e., from the window size and slide. To this end, the operator function is divided into a fragment operator function and an assembly operator function. The fragment operator function defines the processing of a sequence of window fragments and produces window fragment results. The assembly operator function constructs the complete window result by combining the window fragment results. This way, different from classical pane-based splitting, the size of the processed batches is independent of the window slide. 
\subsection{Elasticity}
\label{sec:elasticity}

This section discusses elasticity solutions for SP systems. It starts with centralized approaches where a single controller adapts the parallelism for the complete operator graph. Then, we discuss decentralized approaches where multiple controllers --- each responsible for a sub-set of operators--- adapt the parallelism. Table \ref{tab:ElasticityApproaches} gives an overview of the approaches according to the properties described in Section \ref{sec:elMethods}. A tick marks that an elasticity approach provides the property. These properties are: (1) Does the approach provide real-time guarantees? (2) Does it use a model rather than a threshold (only), (3) Is it a distributed approach? (4) Does it consider state migration?,(5) Does it evaluate its actions to improve its decision making process? Approaches without a tick either exclude these properties or do not consider them further.

\newcolumntype{v}[1]{%
	>{\begin{turn}{90}\begin{minipage}{#1}\raggedright\hspace{0pt}}l%
	<{\end{minipage}\end{turn}}%
	}

\begin{table}[]
	\small
	\begin{tabular}{|l||c|c||c|l|c|c|c|c|c|c||p{1.8cm}|c|}
		\hline 
		Name, Ref & \multicolumn{1}{c|}{ \begin{turn}{90}Data: System \end{turn}}& \multicolumn{1}{c||}{ \begin{turn}{90}Data:Workload \end{turn}} &   \multicolumn{1}{c|}{ \begin{turn}{90}Timing: RE / PR \end{turn}}&  \multicolumn{1}{c|}{ \begin{turn}{90}Objective \end{turn}}&  \multicolumn{1}{c|}{ \begin{turn}{90}Realtime \end{turn}}&  \multicolumn{1}{c|}{ \begin{turn}{90}Scale Up / Out \end{turn}}& \multicolumn{1}{c|}{ \begin{turn}{90}Model-based \end{turn}}  & \multicolumn{1}{c|}{ \begin{turn}{90}Distributed \end{turn}} &\multicolumn{1}{c|}{ \begin{turn}{90}State Migration\hspace{5pt} \end{turn}} &  \multicolumn{1}{c||}{ \begin{turn}{90}Evaluation \end{turn}} & \multicolumn{1}{c|}{ \begin{turn}{90}Infrastructure \end{turn}} &  \multicolumn{1}{c|}{\begin{turn}{90}Year \end{turn}}\\ 
		
		\hline 
		Schneider \cite{schneider_elastic_2009} & x & & RE &TP & & U &  & x &  & x & SM &  2009  \\ 
		\hline 
		Satzger \cite{satzger_esc:_2011} &x & & RE  & Lat	&   & O  &  & &   &  & Cloud &2011 \\ 
		\hline 
		Gulisano \cite{gulisano_streamcloud:_2012} & x & & RE & Res. & & U  &  &  & x &  & Cloud &2012 \\ 
		\hline 
		Tang \cite{tang2013autopipelining} & x & &  RE &TP& & U  &x   &   & x & x &SM &2012 \\ 
		\hline 
		Fernandez \cite{CastroFernandez:2013:ISO:2463676.2465282} & x & x & RE & TP & & O &  &  & x &  & Cloud &2013 \\ 
		\hline 
		Balkesen \cite{balkesen2013adaptive} & x & & PR  &Lat & x & O & x &  &  & &  Cluster&2013 \\ 
		\hline 
	    Akidau \cite{akidau_millwheel:_2013} & x & x& RE & LB, Lat &  & U, O &  &  & x &  &Cluster, Cloud&2013 \\ 
		\hline 
		Gedik \cite{gedik_elastic_2014} & x & & RE & TP &  &  O&  & x  & x & x &Cluster& 2014 \\ 
		\hline 
		Kumbhare \cite{kumbhare_plasticc:_2014} & x & x& PR & TP &  &  O &x  &   &  &  &Cloud & 2014\\ 
		\hline
		Lohrmann \cite{lohrmann_nephele_2014} &  x &  & RE  & Lat, TP & x&  O & x & x &  &  & Cluster&2014 \\ 
		\hline 
		Heinze \cite{heinze2014auto} & x &  & RE & Util, Lat &  &  O & x* & & x & x &Cluster &2014 \\ 
		\hline 
		Heinze \cite{heinze2014latency} &x &  &RE& Migr., Lat	& x & O  & x&  & x & x & Cluster&2014 \\ 
		\hline 
		Heinze \cite{Heinze:2015:OPO:2806777.2806847} & x &  & RE &  Lat & x & O & x  &  & x & x &Cluster &2015 \\ 
		\hline 
		Lohrmann \cite{lohrmann_elastic_2015} &  x & x & PR &  Lat & x & O & x &  &   & x &Cluster& 2015 \\ 
		\hline 
		Mayer \cite{mayer_predictable_2015} &  & x & PR &  Lat& x  & O &x  & x &  &  & Cluster &2015 \\ 
		\hline 
		Zacheilas \cite{zacheilas2015elastic} & & x &PR &Lat & x & O& x&  &  & x & Cloud &2015 \\ 
		\hline
		Sun \cite{SUN201592} &x & &PR &Lat & & O & x & & & &Cloud & 2015\\
		\hline 
		Hochreiner \cite{7820260} & x &  &RE &  Lat &  & O &  & x & &  & Cloud &2016 \\ 
		\hline 
		Hochreiner \cite{7579390} & x & x & RE & Lat & & O & & &x & & Cloud & 2016 \\
		\hline
		Mencagli \cite{Mencagli:2016:GAE:2952298.2903146} &  & x & RE & TP & & O & x & x &  &  &  Cluster&2016 \\ 
		\hline 
		De Matteis \cite{DeMatteis:2016:KCR:2851141.2851148, DEMATTEIS2017302} & x & x & PR & Lat & x & U & x&  & x &  &SM &2016 \\ 
		\hline 
		De Matteis \cite{7912626} & x & x & PR &  Lat &x  & O &x  &  & x &  &Cluster &2017 \\ 
		\hline 
		Hidalgo \cite{HIDALGO2017205} &  x & x & R/P & TP  & & U & x  &  &  & x&Cluster &2017 \\ 
		\hline 
		Kombi \cite{kombi2017preventive} & x & x & PR & TP, Stab.  &  & U, O & x &  & x &  & Cloud &2017 \\ 
		\hline
		Cardellini \cite{cardellini2018decentralized} & x & x &PR & Lat &  & U, O& x* & &  & x & Cloud, Fog& 2018 \\ 
		\hline  
		Mencagli \cite{MENCAGLI2018862} & x & x &PR & LB, Lat &  & U, O & x &  & x &  & Cloud &  2018 \\ 
		\hline 

	\end{tabular} 
	\caption{Categorization of SP operator elasticity literature. The first column contains the first author and the reference. The other columns specify the categorization according to the criteria introduced in Section~\ref{sec:elMethods}. We ordered the table by publication date (last column). Abbreviations: 
		\textit{RE or PR} : reactive or proactive approach;
		    Objectives (Obj.): 
		 \textit{TP}: Throughput, 
		 \textit{Lat}: Latency, 
		 \textit{Res}: Resources, 
		 \textit{LB}: Load Balancing, 
		 \textit{Util}: Resource Utilization, 
		 \textit{Migr}: Number of migrations, 
		 \textit{Stab}: Number of re-configurations; 
		 \textit{RT}: real-time guarantees;
		  \textit{Ev.}: evaluation of scaling effects; 
		  \textit{SM}: Single Machine;  
		  *: Model-based with learning.}
	 
	\label{tab:ElasticityApproaches}
\end{table}

\subsubsection{Centralized Elasticity Solutions}
This section presents elasticity solutions that implement a centralized controller. To further group the approaches, it begins with threshold-based approaches and continues with approaches that rely on a model for their scaling decision.
\paragraph{Threshold-based approaches}
The following approaches reactively scale up or out when a threshold is met. 

Satzger et al.  \cite{satzger_esc:_2011}  propose a distributed SP platform in Erlang called ESC for balanced, fault tolerant elastic stream processing in homogeneous cloud-environments. With this platform, programmers can adapt policies for thresholds, policies for scaling and splitter rules where the thresholds are set for workload and queue length. 
Their platform follows the vision of autonomic computing \cite{1160055}. An ``autonomic manager'' controls the number of VMs and the distribution of operator instances on them.
The contribution of ESC is rather its extensible architecture than the specific methods for elasticity control. 
While operators can be stateful, the authors do not describe a state migration process.
They further leave interference effects of multiple operators on a shared node as future work. 

The elasticity component in VISP by Hochreiner et al.~\cite{7579390} uses thresholds on queue size or delay for each operator. VISP is a distributed SP system specialized on Internet of Things applications. It provides data integration, a ``marketplace'' for operators and operator topologies, and billing. Special is a shared key-value store for state that prevents state migrations. 

With an upper and lower threshold for CPU utilization, in ``StreamCloud" by Gulisano et al. \cite{gulisano_streamcloud:_2012} an ``Elasticity Manager" reactively controls the average CPU utilization of the cluster hosting the operator graph. The authors propose the elasticity management in addition to a new key-based parallelization technique as described in Section \ref{sec:parallelization}. 
Their goal is to minimize resource consumption of an SP system in a private cloud to free resources for other applications.
To scale out fast, a resource manager provides a pool of idle machines that can be activated when needed. 
To manage state, the authors propose an overlap phase and a state migration protocol. The involved operator instances agree on a start time where the former and the now responsible operator instance either start to process the same data until the state can be discarded at the old instance (overlap phase for sliding windows) or migrate state and buffer input data until the migration is finished (migration). 

Similarly,  Fernandez et al. \cite{CastroFernandez:2013:ISO:2463676.2465282} reactively scale out their SP system based on CPU thresholds: Their elasticity mechanism heuristically removes throughput bottlenecks where the average \textit{user and system} CPU utilization exceed a threshold. The system CPU utilization tells about the real load on the machine, possibly induced by other, interfering applications. Again, the solution increases the scale out speed by keeping a ``pool'' of idle VMs. 
 To easily migrate state and be fault tolerant, the implementation frequently stores checkpoints of operator states. 
 Scale in, i.e. merging of states, is mentioned as future work.
 Madsen and Zhou present in \cite{madsen2015dynamic} a similar approach to reduce the latency induced by the state migration with re-using availabe checkpoints. 
 
The MillWheel framework proposed by Akidau et al. \cite{akidau_millwheel:_2013} scales up or out based on thresholds on CPU or memory utilization. The framework provides a fault tolerant, exactly once semantic for highly scalable SP applications and key-based, elastic data parallelization. Consistent state migration is supported with atomic state write operations and tokens that enforce single writer semantics.

\paragraph{Reactive Model-based approaches}
With a greedy algorithm, Tang and Gedik \cite{tang2013autopipelining} dynamically adapt the level of pipelining of an SP system on a multi-core machine. The authors use the assumptions that less utilized threads have a higher throughput and that pipelining and task parallelism are inherently available in the operator graph.
Their solution minimizes the overall utilization for all threads. For fast processing, an exhaustive tree based aggregation reduces the search space of the algorithm. 
To continuously improve, the approach reverses and blacklists adaptations that do not sufficiently improve throughput. To migrate state, the system blocks while spawning new threads. 

The group around Thomas Heinze published three approaches in the topic of elastic scaling. The first compares different auto-scaling strategies. The second tackles the problem of latency spikes due to state migration and the third tackles the question how to find optimal system parameters to configure an elastic SP system. Heinze et al. \cite{heinze2014auto} analyze three different auto-scaling strategies on a cluster: global thresholds, local thresholds and reinforcement learning (RL). The goal is to maximize the utilization of processing nodes (i.e. to reduce the number of required nodes) while keeping the latency low. They use their system FUGU \cite{heinze2013elastic} for evaluations. 
The global threshold strategy scales based on to the average CPU utilization of the cluster, the local based on the CPU utilization per processing node. 
The rewards used in the RL approach depend on the degree the host utilization differs from a target utilization. The authors emphasize the importance to update the RL model according to experienced success or failure. 
The comparison shows that global threshold methodologies are not feasible for fast adaption of SP systems and that the RL-approach leads to the lowest latency with the highest utilization. 

The latency spike a state migration induces is the focus of the follow up work by Heinze et al. \cite{heinze2014latency}. They propose a reactive model-based controller that scales their SP system and guarantees an average end-to-end latency bound while keeping the resource utilization high. Their approach differentiates mandatory and optional migrations. Mandatory are migrations due to node overload (scale out). Optional migrations are due to node underload to improve the overall node utilization (scale in). These optional migrations are carried out only when the expected latency spike due to migration does not violate the latency bound. 
To predict the spike, the algorithm considers the size of the state to migrate, the total numbers of operators to move and the current arrival rate of the system. The latency values used for the cost model are calculated from the expected queue lengths if the system pauses.
To improve the decision making, the system continuously updates its expected pausing times with measured values from performed migrations. 
In case of an optional migration, if releasing a complete host violates the latency bound, operators are migrated stepwise until the host can be shut down.
The authors employ the FLUX-state migration protocol \cite{shah_flux:_2003}. 
Their bin packing algorithm (cf. \cite{heinze2013elastic}) reassigns the selected operators to migrate to new machines based on CPU utilization. 

Finally, Heinze et al.  \cite{Heinze:2015:OPO:2806777.2806847} handle the problem that elasticity strategies require user defined parameters, for example sampling frequencies, thresholds and placement strategies. As these parameters significantly influence the system's performance and cost, they need to be set carefully. The authors propose a model-based parameter-optimization method. With simulations, it automatically tunes six different thresholds to achieve cost-optimality with Heinze et al.'s reactive scaling-method.

\paragraph{Proactive Model-based approaches}
Balkesen et al. \cite{balkesen2013adaptive}  propose a multi-query SP framework with key- and pane-based data parallelization that manages how input streams are assigned to splitter nodes. Their goal is to minimize the number of processing nodes while keeping end-to-end latency bounds and balanced load. With an exponential smoothing technique and user-provided meta-data, they predict the future arrival rate and the change behavior of the input stream.
Given this data, they calculate the number of required splitter nodes and assign them the input streams with a packing algorithm.
Additionally, a mathematical equation is used to calculate the required parallelization degree given the predicted workload, node capacity and processing cost of input data items.
A pool of idle nodes enables fast scale out.
To add an operator instance, splitter and merger stop to connect to the new node. The approach does not include state migrations.

Sun et al.~\cite{SUN201592} propose Re-Stream, a scheduler for distributed and parallel SP systems that is both latency- and energy-aware. A re-scheduling algorithm minimizes the response time of operators on the critical path and schedules the other operators to minimize energy consumption of the system.

Solutions that employ the Kingman's formula from Queuing Theory (QT) \cite{kingman1961single} to calculate parallelization degrees come from Lohrmann et al. \cite{lohrmann_elastic_2015}  and De Matteis and Mencagli  \cite{DeMatteis:2016:KCR:2851141.2851148, DEMATTEIS2017302}. This formula provides average queuing delays for general arrival and processing distributions (G/G/x queues) under heavy load. Worst case analysis with QT (cf. Mayer et al. \cite{mayer_predictable_2015}) is possible for specific distributions only. 

Lohrmann et al. \cite{lohrmann_elastic_2015} extend their system Nephele-streaming  \cite{lohrmann_nephele_2014} we describe in Section \ref{sec:distributedElasticity} with a centralized controller to guarantee an average latency bound while minimizing resource consumption. With Kingman's formula, they predict queuing latencies and adapt the data parallelization degree of the operators in the SP system accordingly. 
To quickly react to sudden workload burst, a reactive component additionally doubles the parallelization degree if an operator becomes a bottleneck. 
The authors assume homogeneous processing nodes. Concerning state migration techniques they refer to other publications in this field. 

De Matteis and Mencagli \cite{DeMatteis:2016:KCR:2851141.2851148, DEMATTEIS2017302} propose latency-aware and energy-efficient scaling of key-based data parallel SP operators. With a Model Predictive Control strategy, they control queuing latencies and energy consumptions on multi-core machines. 
Their goal is to minimize a latency-violation penalty, energy-consumption cost and consider a system stability value to avoid oscillation. Similar to the work of Mayer et al. \cite{mayer_predictable_2015}, a disturbance forecaster predicts the future arrival rate and processing time.
As in Lohrmann et al. \cite{lohrmann_elastic_2015}, De Matteis and Mencagli predict the average per-tuple latency with QT.
Moreover, De Matteis and Mencagli propose a power consumption model that predicts energy consumption for a CPU frequency and number of active cores.
A state migration protocol exchanges state with a shared memory state storage. Special about this migration protocol is that only those instances pause their processing that are involved in the state migration. The other instances continue processing. 
Later, De Matteis and Mencagli extend their approach to horizontal scaling across several machines \cite{7912626}, building on the same predictive control model. 

Zacheilas et al. \cite{zacheilas2015elastic} use a shortest path algorithm to proactively scale applications written with their SP framework (cf. \cite{zygouras2015insights}). Their algorithm minimizes resource cost, penalties for missed tuples due to operator overload and the state migration downtimes. The approach predicts latency and workload using a Gaussian Process and then models possible system configurations as a directed, acyclic graph. A shortest path algorithm finds the cost minimal sequence of configurations. 
 
Hidalgo et al. \cite{HIDALGO2017205} propose a hybrid reactive and proactive elasticity controller implemented on top of the S4 SP system \cite{neumeyer_s4:_2010}.
The elasticity controller has two parts, a reactive short-term adaptation and a proactive mid-term adaptation. It controls the load of an operator based on its workload-to-throughput ratio and aims to maximize the system's throughput, avoiding bottlenecks.  
The short-term adaptation reactively changes the parallelization degree of an operator based on utilization-thresholds. The mid-term adaptation predicts the future load-state of an operator with a Markov-Chain model. It adapts the parallelization degree to the most probable load-state. The authors mention possible state migration solutions that can be included into their approach.

Kombi et al. \cite{kombi2017preventive} published the AUTOSCALE approach that centrally adapts the parallelization degrees of the operators in an operator graph based on predictive input values. Their goal is to proactively avoid congestion within the operator graph but be resource optimal and avoid too frequent reconfigurations. For each operator, AUTOSCALE predicts its future input in two ways: First, using linear regression, it predicts the input event rate of each operator using data from the last monitoring interval. Second, it predicts the input event rate from the predicted output event rate of the upstream operators and its selectivity. To decide about adapting the parallelization degree of a given operator, they combine these two input-estimations and calculate an activity metric from the input and the operators processing capacity. 
The scaling decisions then consider the activity metric and the derivative of the linear regression function where the letter serves as trend-indicator for the input load. While the solution considers stateful operators, the authors do not discuss state migration. 

Cardellini et al. \cite{cardellini2018decentralized} propose a hierarchical controller for a distributed SP system to manage the parallelization degree and placement of operators. Local components send elasticity and migration requests to a global component that prioritizes and approves the requests based on benefit and urgency of the requested action. The cost-metric the global controller minimizes comprises the downtime caused by an action, the performance penalty in case of overloaded operators, and the cost for required resources. The global approval is made using a token bucket implementation.  Regarding elasticity, the authors propose two concepts: A CPU-threshold based and a reinforcement learning based one. The reinforcement concept is further split up in a basic and a model supported solution. The basic solution switches between exploitation and exploration phases. The model supported solution pre-computes possible long-term costs based on the probabilities for parallelization degrees and arrival rates. 
It updates its knowledge at runtime which supersedes an exploitation phase.
Their implementation in D-Storm (cf. \cite{liu2017d}) migrates state with downtimes.
Due to its decentralized nature, the controller is applicable in widely distributed environments.

Mencagli et al.~\cite{MENCAGLI2018862} propose a two-level autonomic adaptation system for pane-based SP operators with two control loops: An inner loop schedules incoming events to worker threads to balance load at bursty data arrival rates. 
An outer loop controls the number of workers for long-term trends in the average data arrival rate. 
The authors argue that the mathematical models for conventional Control Theory methods are too complex given a system with multiple components and interdependent dynamics. Hence, they apply a Fuzzy Logic Controller which a domain expert configures. Several synthetic and real-world scenarios show the Controllers ability to deal with fluctuating workload.
In case of changes in the thread level, the system updates the information that ensures that workers consistently assemble the pane results in the second stage of the pane-based operator.

Most of the elasticity controllers assume that the performance of allocated compute resources is stable. However, Kumbhare et al.~\cite{6846470} observe performance variations in VMs on multi-tenant clouds that require frequent adaptation of the configuration of the SP system (elasticity and placement of operator components). They propose a predictive controller to dynamically re-plan the allocation of elastic cloud resources which mitigates the impact of both resource and workload fluctuations.

\subsubsection{Distributed Elasticity Solutions}
There are a couple of papers that explore \emph{distributed} elasticity controllers in charge of observing and controlling different sub-parts of the operator graph. We group them into approaches where a controller is responsible for multiple operators and approaches that focus on the control of single operators.
\label{sec:distributedElasticity}
\paragraph{Multi-Operator Solutions}
Some solutions find global consensus where the controllers communicate in order to find agreements in their reconfiguration choices. Early work by Weigold et al.~\cite{Weigold:2012:PBI:2330090.2330094} proposes a \emph{rule-based} autonomic controller for distributed components in grid computing. The authors propose an architecture and discuss some implications of their design choices, but do not explicitly state how to configure the controller (i.e., the rules) in order to achieve specific optimization goals. Mencagli \cite{Mencagli:2016:GAE:2952298.2903146} proposes a distributed solution that employs a \emph{game-theoretic} controller for each SP operator. By starting a game, each local controller determines the optimal parallelization degree for its operator, thereby maximizing the throughput while minimizing the monetary cost.
Mencagli compares a non-cooperative and an incentive-based approach that encourages cooperation. The latter leads to a better solution for the whole system than the non-cooperative approach. Additionally, Mencagli et al.~\cite{Mencagli:2014:CPC:2597760.2567929, mencagli2016adaptive} examine distributed elasticity control by applying \emph{Model Predictive Control} in combination with a \emph{cooperative optimization} framework. In particular, the effects of \emph{switching costs} between configurations is modeled by a mathematical function which is used by a proactive control strategy to globally optimize the elasticity decisions. To find agreement between the distributed controllers, they apply the distributed subgradient method to optimize the sum of cost functions over all operators. 

Lohrmann et al. \cite{lohrmann_nephele_2014} propose Nephele-Streaming, an SP framework that uses micro batching of data items. The authors reactively balance throughput and latency, aiming to keep throughput high but keep the average end-to-end latency withing user-defined bounds. They therefore use dynamic output buffer sizes (``adaptive output buffer sizing'') and task fusion (``dynamic task chaining''). The former reduces the output batch size and thus waiting latencies. The latter increases the number of tasks that run within the same thread to reduce communication latencies (``reverse pipelining").  They propose a distributed, model-based approach with a set of QoS managers. Each manager manages the latency constraints for one part of the operator graph based on latency and CPU load measurements. 
When fusing tasks, i.e. omitting queues between them, the system either drops the data items in these queues or waits until the data items in the queues are processed.
Their solution targets big cluster of nodes, which motivates the decentralized approach. 

\paragraph{Single-Operator Solutions} This section summarizes three threshold and one model based approaches that provide elasticity control for single operators (or a limited sequence of operators) only instead of the complete operator graph.
Schneider et al. \cite{schneider_elastic_2009} provide an algorithm that adapts the thread level of a single SP operator to control its  throughput. The algorithm greedily adapts the thread level until throughput is not increased further. Then a stability condition is met and the thread level is kept. If the system detects changes in the throughput, it re-runs the algorithm to adapt the thread level according to changes in the workload as well as higher external load on the system. 
Noteworthy is that the authors provide one global queue the threads share as input queue. They further explicitly consider interferences on the node with other processes and adapt the thread level accordingly.
They assume a single node environment.

Gedik et al. \cite{gedik_elastic_2014}  propose an extension of their key-based data parallelization framework \cite{schneider_auto-parallelizing_2012} that reactively adapts the parallelization degree of parallel regions according to changes in the workload. Their goal is high throughput at low resource usage in a best effort manner.
A threshold based congestion-detection model uses back pressure information to decide about scaling. The splitter detects congestion and measures throughput of its operator. 
To avoid frequent scaling, the elasticity component remembers effective operating states of the system (parallelization degree) for the current workload. When the workload changes, the remembered system states are purged, and the system settles again to the new workload. A level function defines how many instances to add or remove. Setting the level function can make the adaptation more or less aggressive. To migrate state, instances store it in a database to make it available to newly spawned operator instances. During migration, the splitter stops. 
The proposed consistent hash function minimizes required state migration and ensures load balancing.
To apply this solution in a multi-operator graph, one provides one splitter per parallel region (cf. \cite{schneider_auto-parallelizing_2012}, Section \ref{sec:key-basedSystems}).

Hochreiner et al. \cite{7820260} propose a reactive elasticity controller for their distributed SP platform PESP. For each operator, the controller minimizes \textit{monetary} costs while ensuring moderate queuing times and CPU utilization with threshold driven scaling. The optimization algorithm considers real cloud provider pricing models that include a minimum time a VM needs to be rented. If an instance shall be shut down due to over provision, the controller selects the instance that has the least left time already paid for.
With state being stored in a shared directory, operator scaling does not require state migration.

A proactive, model-based approach comes from Mayer et al. \cite{mayer_predictable_2015}. They provide a \textit{window}-based data parallelization single operator framework. The operator instances can thereby be distributed to multiple machines.  An elasticity controller proactively ensures a \textit{worst case} latency limit using a queuing-theory based model. Thereby, all operator instances are assumed to behave homogeneously, i.e. have the same service times. While this model is limited to specific probabilistic distributions of arrival and processing rates, it enables the worst case limitation of queuing latency. 
In their framework, windows, when assigned to an operator instance once, will not be migrated.
If an instance shall be shut down, it finishes the processing of its assigned windows and shuts down afterwards. Hence, scaling does not require state migration or downtime. 
\section{Related Work}
\label{sec:related}

While this survey focusses on SP operator parallelization and elasticity, there are further important aspects when it comes to manage SP systems. This section provides a brief overview and discussion. The following aspects are discussed:  Placement of SP operators,  efficient resource provision, and efficient processing algorithms for SP operators.  Furthermore, we discuss the relation of SP systems to batch processing systems.

\subsection{Operator Placement}
\label{sec:placement}

One important aspect in SP system management is the placement of SP operators in a distributed infrastructure. In the realm of SP system, this problem is often referred to as the \emph{scheduling problem}, i.e., where to execute which parts of the operator graph. This regards the coarse-plained operator placement in different domains, e.g., in different data centers, but also the fine-grained placement within a single domain, e.g., within a single data center. Due to dynamic workloads of an SP system, operator placement cannot be a static decision, but operators need to be frequently migrated. In this section, we discuss some general challenges and solutions to this end. 

Cardellini et al. \cite{Cardellini:2016:OOP:2933267.2933312} formulate the placement problem as an integer linear program (ILP) that can be solved optimally with an ILP solver. However, due to the NP-completeness of the problem, there is the need for efficient heuristics. SBON by Pietzuch et al. \cite{pietzuch_network-aware_2006} is a placement algorithm that is based on a spring relaxation model and optimized the network utilization. Rizou et al. \cite{5560127} propose another placement algorithm that uses the gradient descent method and optimizes the network utilization. SODA \cite{Wolf:2008:SOS:1496950.1496970} by Wolf et al. is an early scheduler proposed for IBM's System S. 
To neither overload the network nor processing capacities of the computing nodes, SODA periodically admits or rejects new operator graphs and places them optimally. Amini et al. \cite{1648858} propose a two-tiered scheduler that maximizes system throughput. The first tier decides about long-term operator placement, while the second tier reacts to bursts in workload with short-term CPU scheduling and event flow control on each computing node. 
When sources and sinks in the operator graph are mobile, the optimal operator placement changes over time, so that operators need to be migrated. To this end, MigCEP \cite{Ottenwalder:2013:MOM:2488222.2488265} by Ottenw\"{a}lder et al.  performs a plan-based migration of operators by predicting mobility patterns. Operator migration can in some cases be too expensive or infeasible. Xing et al. \cite{Xing:2006:PRL:1182635.1164194} have developed a placement algorithm that provides \emph{resilient} placements that can withstand workload changes without operator migrations. Similarly, Drougas and Kalogeraki~\cite{5161015} propose a method for operator placement and parallelization that is resilient to sudden \emph{bursts} in the input streams. 

With the increasing need for operator parallelization and elasticity, fine-grained placement, i.e., where to place which operator components, has become an urgent problem. Aniello et al. \cite{Aniello:2013:AOS:2488222.2488267} propose an adaptive online scheduler tailored to the Storm ESP system that takes into account traffic patterns between components of the ESP system to reduce inter-node traffic. The T-Storm scheduler by Xu et al. \cite{6888929} follows a similar goal; in comparison to Aniello et al. \cite{Aniello:2013:AOS:2488222.2488267}, T-Storm introduces a couple of further optimizations tailored to the Storm SP system.  Fischer and Bernstein \cite{7363749} model the communication between the instances in Storm as a graph and solve a graph partitioning problem to minimize the communication while keeping the computational load between the nodes balanced. Their approach shows a better performance than the aforementioned adaptive online scheduler by Aniello et al. \cite{Aniello:2013:AOS:2488222.2488267}.

There are also combined approaches which regard determining the operator parallelization degree and the fine-grained placement of operator instances as a holistic problem. Cardellini et al. \cite{Cardellini:2017:OOR:3092819.3092823} formalized the combined problem for data-parallel SP as an integer linear program (ILP) that can be solved optimally with an ILP solver. 
P-Deployer \cite{7962193} by Liu and Buyya models the same problem as a bin-packing problem and solves it by a heuristic that is based on the first fit decreasing method.  Backman et al. \cite{backman_managing_2012} propose a scheduling framework that can exploit data and task parallelism in SP operators. It minimizes the end-to-end latency with operator parallelism and the scheduling of the corresponding operator components on the available computing node. The authors use a tiered bin-packing problem, which allows for prioritizing operators, and solve the optimization problem using simulation-based latency estimation. Madsen et al. \cite{madsen_integrative_2017}  provide a solution to integrate load balancing, collocating (i.e., placement), and horizontal scaling (i.e., determining the parallelization degree of operators). They model the problem as a Mixed-Integer Linear Program, solved with a heuristic greedy algorithm. While the papers discussed in this paragraph have some overlap with the work on operator elasticity discussed in Section \ref{sec:elasticity} their focus is more on the placement problem than on elasticity control.

\subsection{Efficient Resource Provisioning}
\label{sec:efficientScaling}

Another aspect that is related both to parallelization and elasticity is how to efficiently provide the required resources in cloud environments.

The SP system Stela by Xu et al. \cite{7484160} tackles the problem which operator to scale out when there are more resources added to the system, and which operators to scale in when resources are removed. Stela maximizes the post-scaling throughput by scaling out the operators that have the highest predicted impact on the application throughput based on an analysis of the congestion in the operator graph. 

Borkowski et al. \cite{borkowski_moderated_nodate} tackle the problem of minimizing the number of superfluous scaling activities at highly noisy workloads. To this end, they employ non-linear filtering techniques from the field of signal processing.

Lombardi et al.  \cite{8067517} make the observation that the scaling of operators and the scaling of resources are two independent tasks that do not necessarily have to be performed jointly. They propose the ELYSIUM controller that first adapts the parallelization degree for each operator, and then adapts the resource provisioning only when needed. To this end, they leverage a resource estimator that predicts the resource consumption based on the current resource utilization. Similarly, Van der Veen et al. \cite{7184876} propose a controller to automatically adjust the number of virtual machines assigned to a deployment of the Storm ESP system.

\subsection{Efficient Operator Execution and Re-Use}
\label{sec:efficientExecution}

Besides questions of scaling and placement of SP operators, the efficient execution of operators and whole operator graphs has received a lot of attention in the literature. On the one-hand side, the execution of single operators is optimized by efficient processing algorithms. On the other-hand side, the execution of overall operator graphs is optimized by operator re-use among multiple concurrent queries.

ZStream by Mei et al.  \cite{mei_zstream:_2009} implements a tree-based pattern-plan structure and dynamically finds the optimal plan to evaluate the CEP-pattern. 

The work of Poppe et al. \cite{Poppe:2017:CET:3035918.3035947} manages the trade-off between memory consumption and processing throughput when detecting sequences of arbitrary, statically unknown length, e.g. Kleene-closure queries. The approach stores common subsequences of multiple pattern instances for re-use, to decreasing the processing load at the cost of memory consumption.

There is a large body of work on the efficient incremental processing of general queries on sliding windows in SP operators. The tutorial by Hirzel et al. \cite{Hirzel:2017:SAA:3093742.3095107} provides a comprehensive overview. Tangwongsan et al. in \cite{tangwongsan_general_2015} and \cite{Tangwongsan:2017:LSA:3093742.3093925} optimize processing depending on the mathematical properties of the operator function (invertability, associativity and commutativity). Le-Phuoc et al. \cite{Le-Phuoc:2013:ESP:2717129.2717148} propose an incrementally sliding window approach for the parallel processing of multiway join and aggregation operations.

When there are multiple queries in an SP system, re-use of results from one query in another query is a common method to reduce overhead. The RECEP system by Ottenw\"{a}lder et al. \cite{Ottenwalder:2014:RSR:2611286.2611297} tackles multi-query SP systems in mobile scenarios, where there are many similar queries that show overlapping ranges of interest in time and space. RECEP allows to re-use results of similar queries, guaranteeing a user-defined quality requirement with precision and recal. SQPR by Kalyvianaki et al. \cite{5767851} and SPASS by Ray et al. \cite{ray_scalable_2016} are placement optimizers that leverage the sharing of computations between the sub-patterns of multiple queries.  SlickDeque by Shein et al.~\cite{Shein2018SlickDequeHT} improves throughput and latency of incremental invertible and noninvertible SP operators and supports efficient multi-query processing.

Verner et al.~\cite{Verner:2011:PDS:1995896.1995915} propose a deadline-aware scheduler for hybrid compute platforms for operators that process multiple different input streams in a highly-parallel fashion. Their solution consist of multiple CPUs and a single accelerator (such as a GPU).

\subsection{Batch Processing Systems}
\label{sec:batching}

Modern batch processing systems make heavy use of data parallelism. Those systems are often based on the MapReduce paradigm presented by Dean and Ghemawat \cite{dean_mapreduce:_2004}, that splits the input data with keys. A couple of works push the traditional MapReduce more toward a streaming behavior. In \cite{lam_muppet:_2012}, Lam et al. present Muppet, a MapReduce modification where an ``update'' function" replaces the reduce function. The update function makes intermediate results accessible any time, giving the impression of a results stream. Similarly, Hadoop Online by Condie et al. \cite{condie_mapreduce_2010, condie_online_2010} allows users to see ``early returns'', i.e., intermediate results of the reduce function. Stream MapReduce by Brito et al. \cite{brito_scalable_2011} introduces ``windowed reducers'' to output a stream of results according to a window policy. Kumbhare et al.~\cite{7164919} extend Stream MapReduce by methods for adaptive load-balancing, runtime elasticity and fault tolerance. Beyond approaches to adapt a streaming model in MapReduce, Apache Flink \cite{Carbone2015ApacheFS}, Apache Spark \cite{Zaharia:2013:DSF:2517349.2522737} and AJIRA \cite{6888930} support batch and stream processing.

\section{Conclusion and Outlook}
\label{sec:conclusion}

As the rate of publications on parallelization and elasticity in SP is still increasing, we see the demand for a structured overview of this field. In this survey, we discussed and classified solutions for parallelization and elasticity in SP systems to provide a comprehensive overview. Besides providing an overview, with our work we hope to enhance the mutual understanding of research communities that look at SP systems from different angles. For instance, we noticed that researchers from the general SP domain typically assume SP operators that have key-partitioned state. In contrast, researchers from the CEP domain focus on window-based operations. Hence, the solutions for parallelization and elasticity that are proposed from the different domains are different. 
For future research in parallel and elastic SP, we observe a couple of recent trends that will impact SP systems. These general trends are fog and edge computing, in-network computing (e.g., on smart NICs), and sophisticated cloud cost models.

As we found in this survey, most of the current solutions on SP parallelization and elasticity are developed for homogeneous resources, often provided by a cloud data center. As a future work, we encourage an extension towards heterogeneous resources, especially considering the upcoming trend toward fog and edge computing that comes with more heterogeneous processing nodes~\cite{Bonomi:2012:FCR:2342509.2342513}. Further, fog and edge compute nodes are limited in their computational capabilities, so that the ``illusion of unlimited hardware'' provided by cloud computing may not hold at the edge. Here, load shedding or approximate computing may be unavoidable, and we see first methods for approximate SP being proposed~\cite{Quoc:2017:SAC:3135974.3135989}.

We see an emerging field of new programmable network devices that allow for offloading computing from CPUs to the network, along with network programming languages such as P4. Early work on in-network CEP points to tremendous potential of in-network computing for this domain, but also reveals challenges due to limitations both in the programmable hardware as well as in the network programming language~\cite{Kohler:2018:PTI:3229591.3229593}.

Finally, promising QoS-metrics to consider in elasticity approaches are energy consumption to support environmental friendly IT solutions and the cost models of cloud providers that go beyond the standard "pay-as-you-go" model. These models support, e.g, "spot instances" and thus make room for financial savings \cite{zheng_how_2015}.

%

	\bibliographystyle{ACM-Reference-Format}
	\bibliography{references}

\end{document}